\documentclass[12pt]{iopart}

\newcommand{\be}{\begin{eqnarray}}
\newcommand{\ee}{\end{eqnarray}}
\newcommand{\bea}{\begin{eqnarray}}
\newcommand{\eea}{\end{eqnarray}}

\usepackage{graphicx}
\usepackage{bm}
\usepackage{color}
\usepackage{slashed}
\usepackage{cite}
\usepackage{graphicx}
\usepackage{multicol}
\usepackage{graphicx}
\usepackage{color}
\usepackage{slashed}
\usepackage{CJKutf8}
\begin{document}
\begin{CJK}{UTF8}{<font>}
\title{Optical appearance of the Schwarzschild black hole in the string cloud context}

\author{Sen Guo$^{1}$, \ Yu-Xiang Huang$^{2}$, \ Guo-Ping Li$^{*2}$}

\address{$^1$Guangxi Key Laboratory for Relativistic Astrophysics, School of Physical Science and Technology, Guangxi University, Nanning 530004, People's Republic of China}
\address{$^2$School of Physics and Astronomy, China West Normal University, Nanchong 637000, People's Republic of China}

\ead{sguophys@126.com; yxhuangphys@126.com; gpliphys@yeah.net}
\vspace{10pt}
\begin{indented}
\item[]Apr. 2023
\end{indented}

\begin{abstract}
The image of a black hole (BH) consists of direct and secondary images that depend on the observer position. We investigate the optical appearance of a Schwarzschild BH in the context of a string cloud to reveal how the BH's observable characteristics are influenced by the inclination angle, string cloud parameter, and impact parameter. Following Luminet's work [Astron. Astrophys. 75, 228 (1979)], we adopt a semi-analytic method to calculate the total bending angle of the light ray and derive the direct and secondary images of the Schwarzschild string cloud BH. Our results show that an increase in the inclination angle leads to a more pronounced separation of the images. We consider the gravitational redshift and present the redshift distribution of the direct image while illustrating the flux distribution. We observe that the direct image exhibits blueshift and redshift simultaneously, and the asymmetry of the flux distribution increases with the inclination angle. Finally, we obtain the Schwarzschild string cloud BH image via a numerical simulation, which provides an approximate illustration of the EHT resolution.
\end{abstract}


\section{Introduction}
\label{intro}
\par
There is compelling astronomical evidence supporting the existence of black holes (BHs) in the universe. Laser-Interferometer Gravitational Wave-Observatory (LIGO) and Virgo detectors captured $\sim 100$ instances of gravitational wave signals from the merging of compact objects \cite{1}. However, the event horizon, which is a defining feature of BHs, can only be indirectly inferred. The Event Horizon Telescope (EHT) captured an image of a supermassive BH surrounded by an accretion flow in the Messier (M) 87$^{*}$ elliptical galaxy \cite{2}. This image revealed that the space around the BH is illuminated by a magnetically arrested accretion disk \cite{3}. Strikingly, the first horizon-scale radio observations of Sagittarius (Sgr) A$^{*}$ in the center of the Milky Way confirmed that the measured ring size of Sgr A$^{*}$ is consistent with the shadow critical curve predicted by general relativity (GR) within 10$\%$ \cite{4}. Taken together, these astronomical observations strongly suggest that supermassive BHs are typically accompanied by luminous accretion disks.

\par
It is well-known that the active galactic nucleus (AGN) is primarily composed of an astrophysical BH and a luminous accretion disk. The observation of the AGN is dependent on the BH and accretion disk, and the appearance of the BH and disk is affected by the spacetime structure and observation inclination. Theoretically, the image of the spherically symmetric BH is a standard circle \cite{5}, while a rotating BH has a D-shape appearance owing to the dragging of lightlike geodesics by the BH's angular momentum \cite{6}. Falcke $et~al.$ proposed ray-tracing codes utilizing relativistic numerical simulations, suggesting that the BH shadow could be observable \cite{7}. In addition to researching BH shadow characteristics within the framework of GR, exploring the image of the accretion disk around a BH is an intriguing topic. The first standard accretion disk model, which presented the physical characteristics of a geometrically thin and optically thick disk, was established in 1973 \cite{8}. Luminet investigated the image of the accretion disk under the Schwarzschild BH background and obtained direct and secondary images of an accretion disk using the semi-analytic method \cite{9}. Since then, many researchers have discussed this topic \cite{10,11,12,13}. Utilizing numerical simulations for ray-tracing and radiative transfer, an elegant description of the BH shadow and rings was proposed by considering a thin accretion disk surrounding a Schwarzschild BH \cite{14}. For an optically thin and geometrically thick disk accretion, Cunha $et~al.$ studied the image of the Schwarzschild BH and showed that an almost equatorial observer can observe different patches of the sky near the equatorial plane \cite{15}. The description of the accretion disk under various modified gravity frameworks can be found in \cite{16,17,18,19,20,21,22,23}.

\par
The physical concept of string theory posits that the fundamental components of nature are one-dimensional strings, rather than zero-dimensional particles. These strings exhibit linelike, cylindrical, and axially symmetric structures, which arise as topological defects resulting from symmetry breaking \cite{24}. String clouds are collections of strings that formed during the early stages of the universe's development as a result of this symmetry breaking. The existence of such strings in the early universe is consistent with observations, and their effects must be taken into account on both gravitational and cosmological scales. When the cosmic temperature was lower than a critical value, the symmetry was broken, and energy became trapped in a specific area of space. Letelier examined the physical characteristics of the string cloud model and obtained a Schwarzschild string cloud BH solution \cite{25}. Recently, researchers have extensively studied various BH properties within the string cloud framework, including thermodynamic behavior \cite{26,27,28,29}, BH solutions \cite{30,31,32}, BH shadows, and gravitational lensing \cite{33,34,35,36,37}. In this article, we only discuss the effect of strong deflection gravitational lensing and do not investigate weak deflection. As future directions, considerable research has been conducted on weak deflection gravitational lensing to probe the Schwarzschild BH in the string cloud \cite{59,60,61,62,63}, as well as the geodesic motion of a particle beyond the innermost stable orbits \cite{64,65,66,67,68,69,70,71,72,73,74}. However, a detailed investigation of these topics is beyond the scope of this paper and will be left for our next study.

\par
The optical appearance of the Schwarzschild BH surrounding a accretion disk in the string cloud context remains an open question. This paper aims to address this issue by using a ray-tracing code and semi-analytic method to extend the study of Schwarzschild BH accretion disk images to the string cloud background. Specifically, we investigate the direct and secondary images of the accretion disk and analyze the impact of the string cloud on the optical appearance of the BH, taking into account the distribution of redshift and flux of the images. The remainder of this paper is structured as follows. Section \ref{sec:2} provides a brief overview of the Schwarzschild string cloud BH and discusses the effective potential. Section \ref{sec:3} presents the derivation of the direct and secondary images of the target BH. In Section \ref{sec:4}, we present the distribution of radiation flux and redshift of the thin accretion disk for various observation angles. Finally, we conclude the paper in section \ref{sec:5}.

\section{Schwarzschild string cloud BH and effective potential}
\label{sec:2}
\par
The metric of the Schwarzschild BH in the string cloud context is given by \cite{25}
\begin{equation}
\label{2-1}
{\rm d}s^{2}=-f(r){\rm d}t^{2}+\frac{1}{f(r)}{\rm d}r^{2}+r^{2}{\rm d}\theta^{2}+r^{2}\sin^{2}\theta {\rm d}\phi^{2},
\end{equation}
where $f(r)$ is the metric potential, which is given as
\begin{equation}
\label{2-2}
f(r)=1-a-\frac{2M}{r},
\end{equation}
where $M$ represents the BH mass, and $a$ is a parameter describing the string cloud. It is worth noting that as $a$ approaches zero, the BH degenerates into a Schwarzschild BH. Therefore, the analysis also includes the scenario of a Schwarzschild BH. Using Eq. (\ref{2-2}), we can express the radius of the BH horizon as follows:
\begin{equation}
\label{2-3}
r_{\rm H}=\frac{2M}{1-a}.
\end{equation}
The string cloud model can provide an explanation for the field theory resulting from the distance interaction between particles, which is associated with a unique behavior of the gravitational field. Assuming that the gravitational field can be created by the elements of the string, we concentrate on the scenario where $a<1$, as it behaves as an attractive gravitational charge \cite{35}.

\par
The motion of photons satisfies the Euler-Lagrangian equation
\begin{equation}
\label{2-4}
\frac{{\rm d}}{{\rm d}\lambda}\Bigg(\frac{\partial {\rm \mathcal{L}}}{\partial \dot{x}^{\rm \alpha}}\Bigg) = \frac{\partial {\rm \mathcal{L}}}{\partial x^{\rm \alpha}},
\end{equation}
where $\lambda$ is the affine parameter, and $\dot{x}^{\alpha}$ is the four-velocity of the photon. We only consider the photons that move on the equatorial plane; hence, the Lagrangian equation is
\begin{equation}
\label{2-5}
\mathcal{L}=-\frac{1}{2}g_{\rm \alpha \beta}\frac{{\rm d} x^{\rm \alpha}}{{\rm d} \lambda}\frac{{\rm d} x^{\beta}}{{\rm d} \lambda}=0.
\end{equation}
The energy and the angular momentum of the photons are conserved quantities, i.e.,
\begin{equation}
\label{2-6}
E=-g_{\rm t \rm t}\frac{{\rm d} t}{{\rm d} \lambda} = f(r)\frac{{\rm d} t}{{\rm d} \lambda},~~~~L=g_{\rm \phi \rm \phi}\frac{{\rm d} \phi}{{\rm d} \lambda}=r^{2} \frac{{\rm d} \phi}{{\rm d} \lambda}.
\end{equation}
The four-velocity of the time, the azimuthal angle, and the radial components can be obtained by using Eqs. (\ref{2-2})-(\ref{2-5}); we have
\begin{eqnarray}
\label{2-7}
&&\frac{{\rm d} t}{{\rm d} \lambda}=\frac{1}{b}\Bigg(1- a - \frac{2M}{r}\Bigg)^{-1},~~~\frac{{\rm d} \phi}{{\rm d} \lambda}=\frac{1}{r^{2}},\\
\label{2-8}
&&\frac{{\rm d} r}{{\rm d} \lambda}=\pm\sqrt{\frac{1}{b^{2}}-\frac{1}{r^{2}}\Bigg(1- a - \frac{2M}{r}\Bigg)}.
\end{eqnarray}
Here, the symbol ``$\pm$" indicates that the photons move radially outward ($-$) or inward ($+$), and $b$ is the impact parameter, which is defined as $b\equiv L/E$. Therefore, the effective potential of the Schwarzschild string cloud BH is
\begin{equation}
\label{2-9}
{\rm V}_{\rm eff}=\frac{1}{r^{2}}\Bigg(1- a - \frac{2M}{r}\Bigg).
\end{equation}
The radius of the photon ring can be obtained from the relations ${\rm V}_{\rm eff}=\frac{1}{b_{\rm c}}$ and ${\rm V}_{\rm eff}'=0$. Thus, the radius of the photon sphere is $r_{\rm s}=\frac{3M}{1-a}$, and the critical impact parameter is $b_{\rm c}=\frac{3\sqrt{3}M}{\sqrt{1-3a+3a^{2}-a^{3}}}$. According to Eqs. (\ref{2-8}) and (\ref{2-9}), we have
\begin{equation}
\label{2-10}
\Omega(u) \equiv \Bigg(\frac{{\rm d} u}{{\rm d} \phi}\Bigg)^{2} = 2 M u^{3} + au^{2} - u^{2} + \frac{1}{b^{2}},
\end{equation}
where $u$ is defined as $u \equiv 1/r$. If $r>r_{s}$, both photons with $b>b_{c}$ and $b<b_{c}$ can reach the distant observer at infinity. Therefore, in the case where $b>b_{c}$, we only consider the light that can reach the observer's plane. This is because when $b<b_{c}$, the light enters the BH and ultimately falls into the singularity. According the Cardano formula, Eq. (\ref{2-10}) can be rewritten as
\begin{equation}
\label{2-11}
2Mu^{3}+au^{2}-u^{2}+\frac{1}{b^{2}} \equiv 2 M G(u)= 2M (u-u_{1})(u-u_{2})(u-u_{3}).
\end{equation}
Here, the cubic polynomial $G(u)$ has two positive roots and one negative root, satisfying $u_{1} \leq 0 < u_{2} < u_{3}$. Following Ref. \cite{9}, these three roots can be given by the periastron distance $P$. By introducing a parameter $Q^{2} \equiv (P-2M)(P+6M)$, one can obtain
\begin{equation}
\label{2-12}
u_{1}=\frac{P-2M-Q}{4MP},~~~u_{2}=\frac{1}{P},~~~u_{3}=\frac{P-2M+Q}{4MP},
\end{equation}
and
\begin{equation}
\label{2-13}
b^{2}=\frac{1}{\frac{P-2M}{P^{3}} - a u^{2}}.
\end{equation}
Hence, we can obtain the value of parameter $b$ at infinity for a given periastron distance $P$, and vice versa. Furthermore, the bending angle of the light ray is
\begin{equation}
\label{2-20}
\psi (u) = \sqrt{\frac{2}{M}} \int_{0}^{u_{2}} \frac{{\rm d}u}{\sqrt{(u-u_{1})(u-u_{2})(u-u_{3})}} - \pi.
\end{equation}
By converting the above equation into an elliptic integral, one obtains
\begin{equation}
\label{2-21}
\psi (u) = \sqrt{\frac{2}{M}} \Bigg(\frac{2 F(\Psi_{1},k)}{\sqrt{u_{3}-u_{1}}} - \frac{2 F(\Psi_{2},k)}{\sqrt{u_{3}-u_{1}}}\Bigg) - \pi,
\end{equation}
where $\Psi_{1}=\frac{\pi}{2}$, $\Psi_{2}=\sin^{-1}\sqrt{\frac{-u_{1}}{u_{2}-u_{1}}}$, and $k=\sqrt{\frac{u_{2}-u_{1}}{u_{3}-u_{1}}}$. According to Eq. (\ref{2-12}), the total change of the bending angle is
\begin{equation}
\label{2-22}
\psi (u) = 2 \sqrt{\frac{P}{Q}} \Big(K(k)- F(\Psi_{2},k)\Big) - \pi,
\end{equation}
where $K(k)$ is the complete elliptic integral of the first kind. Note that the above calculation is based on Luminet's semi-analytic method for calculating the total bending angle of the light ray \cite{9}. However, there are other methods for calculating the bending angle in strong fields, particularly the strong deflection limit proposed by Bozza \cite{38}. In this limit, the deflection angle can be determined as follows:
\begin{equation}
\label{2-14}
\phi (x) =  - \overline a \log \Big(\frac{{\theta {D_{\rm OL}}}}{{{b_{\rm c}}}} - 1 \Big) + \overline b.
\end{equation}
Here, $\overline a$ and $\overline b$ are the strong deflection limit coefficients, which satisfy
\begin{equation}
\label{2-15}
\overline a  = \frac{{R(0,{x_{\rm m}})}}{{2\sqrt {{\beta x_{\rm m}}} }},~~~~~~\overline b  =  - \pi  + {b_{\rm R}} + \overline a \log \frac{{2{\beta_{\rm m}}}}{{{y_{\rm m}}}},
\end{equation}
where the subscript $m$ denotes the value at $x = x_{\rm m}$, and
\begin{equation}
\label{2-16}
{b_{\rm R}} = \int_{0}^{1} g(z,{x_{\rm m}}){\rm d}z,
\end{equation}
\begin{equation}
\label{2-17}
R(z,{x_{\rm m}}) = \frac{2 \sqrt{B y}}{Cd(A)} (1-y_{\rm m}) \sqrt{C_{\rm m}}
\end{equation}
\begin{equation}
\label{aaa}
g(z,x_{\rm m}) = R(z,x_{\rm m})f(z,x_{\rm m}) - R(0,x_{\rm m})f_{\rm m}(z,x_{\rm m}),
\end{equation}
\begin{equation}
\label{2-18}
f(z,x_{\rm m}) = \frac{1}{\sqrt{y_{\rm m}}-\big((1-y_{\rm m})z + y_{\rm m}\big)\frac{C_{0}}{C}},
\end{equation}
in which $A(x)$, $B(x)$, the $C(x)$ are obtained from $f(x)$ deformation; we have
\begin{equation}
\label{2-19}
A(x) = B(x)^{-1} = 1 - \frac{1}{r} - a,~~~C(x) = x^2,~~~y = A(x),~~~z =\frac{y-y_{\rm m}}{1 - y_{\rm m}}.
\end{equation}
The researchers extensively investigated the bending angle of light rays in strong gravitational fields, by using various modified gravity contexts as a means of obtaining the angle. This is the second method they employed \cite{39,40,41,42,43,44,45,46,47,48,49,50,51,52,53,54,55,56}.
\begin{center}
\includegraphics[width=5cm,height=5cm]{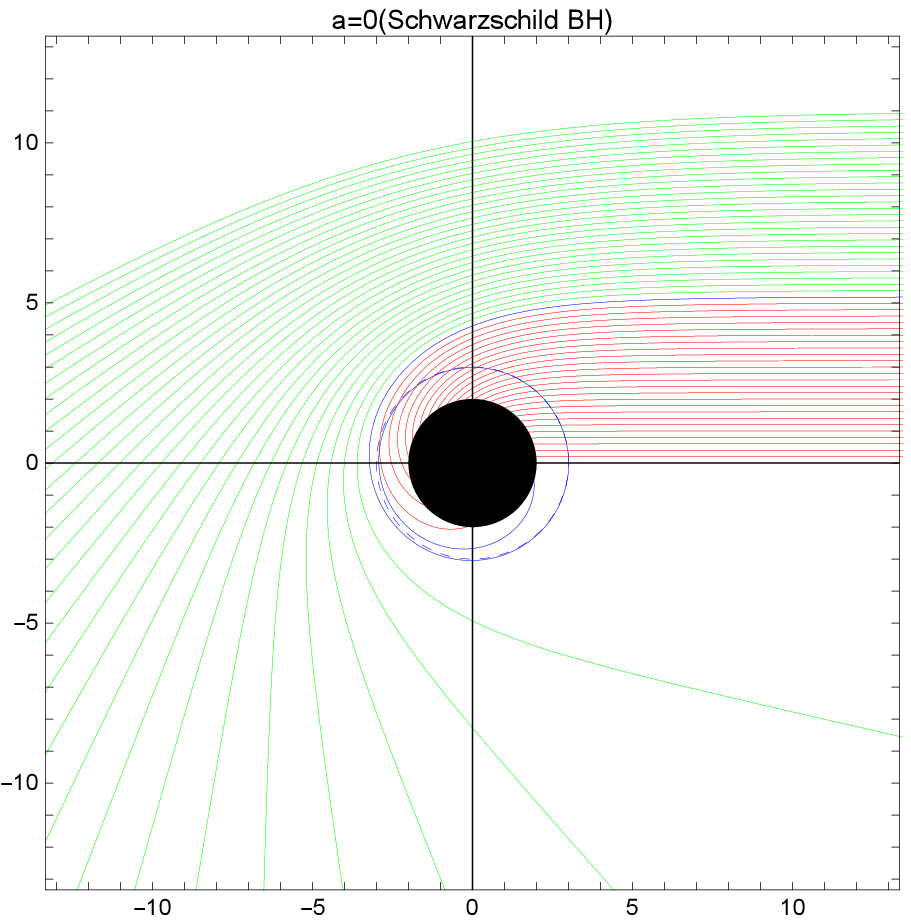}
\includegraphics[width=5cm,height=5cm]{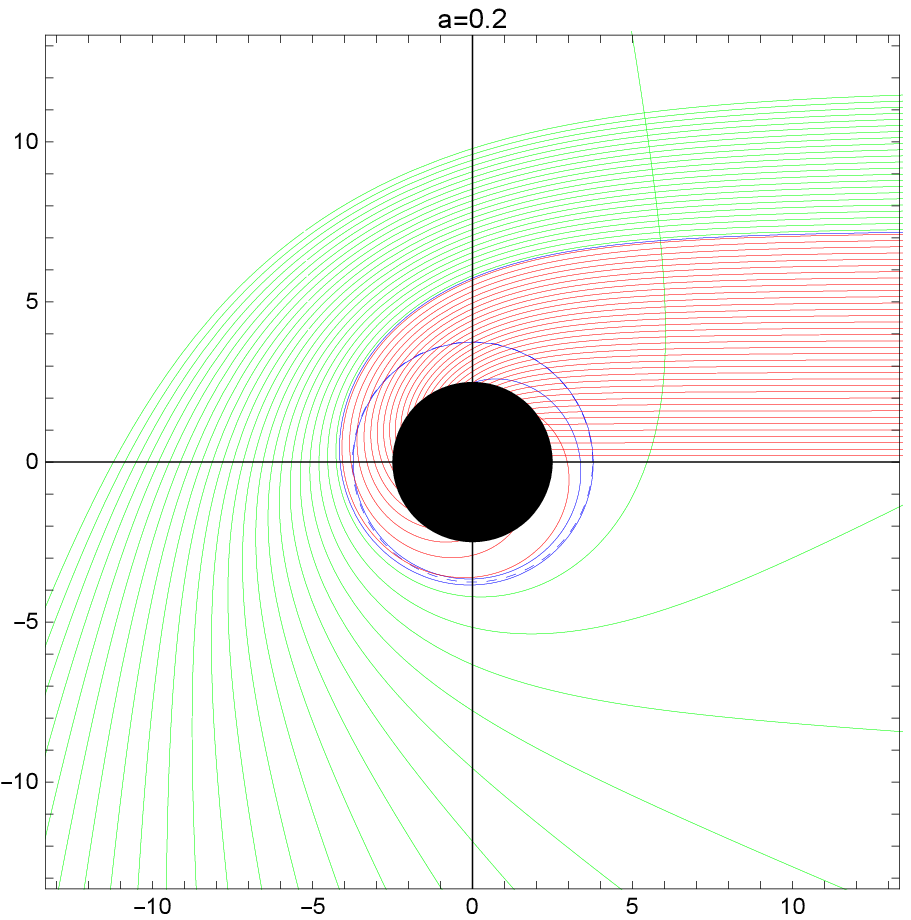}
\includegraphics[width=5cm,height=5cm]{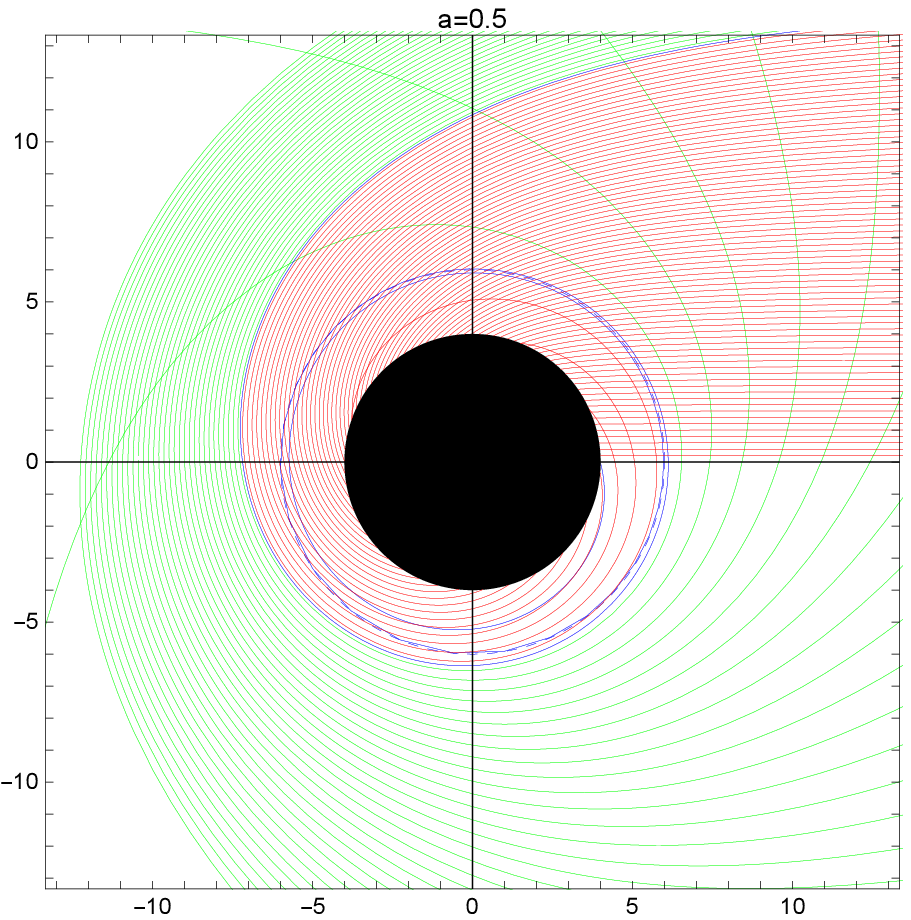}
\parbox[c]{15.0cm}{\footnotesize{\bf Fig~1.}  
The trajectory of the light ray for different $a$ values with $M=1$ in polar coordinates $(r,\phi)$. {\em Left Panel}-- string cloud parameter $a=0$ (Schwarzschild BH), {\em Middle Panel}-- string cloud parameter $a=0.2$, and {\em Right Panel}-- magnetic charge $a=0.5$. The green, blue, and red lines correspond to $b>b_{\rm c}$, $b=b_{\rm c}$, and $b<b_{\rm c}$, respectively. The BH is shown as a black disk.}
\label{fig1}
\end{center}
\par
By using a ray-tracing code, we can observe the trajectory of a light ray under different string cloud parameters, as illustrated in Fig. 1 through the bending angle function. It is evident from the figure that a larger value of $a$ corresponds to a larger radius of the black disk. Light rays in proximity to the BH with a large $a$ may be extremely curved, resulting in an increase in the light ray density for a distant observer.

\section{Direct and secondary images of Schwarzschild string cloud BH}
\label{sec:3}
\par
In this section, we employ the Luminet method to examine the direct and secondary images of the Schwarzschild string cloud BH. Our assumption is that the black hole is surrounded by an optically thick and geometrically thin accretion disk. The radiation is emitted from point $M$ (the emission plane) with coordinates $(r,\varphi)$ and travels to point $m$ (the observation plane) with coordinates $(b,\alpha)$. The coordinate system of the accretion disk can be seen in Fig. 2.
\begin{center}
\includegraphics[width=6.5cm,height=5cm]{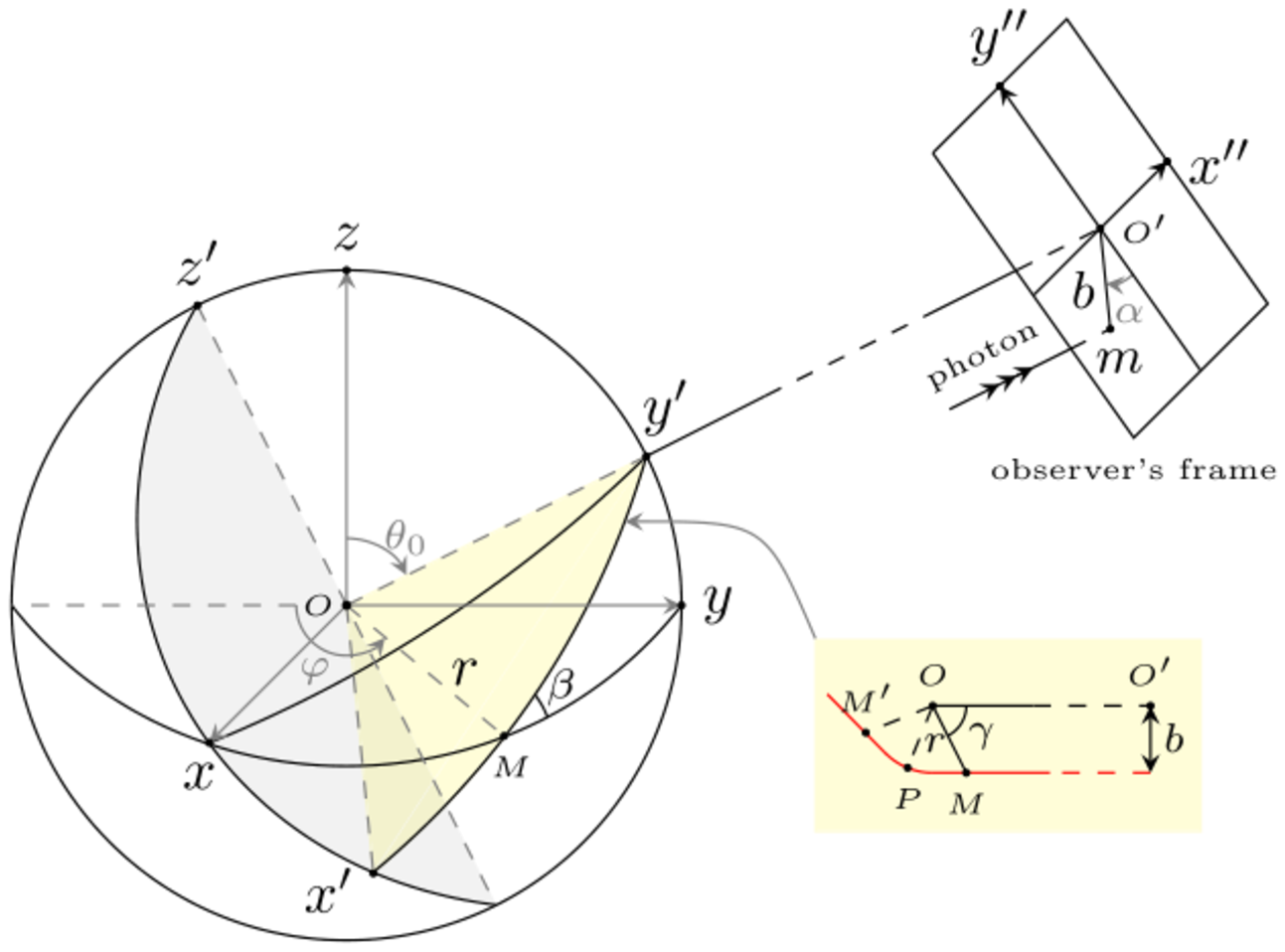}
\parbox[c]{15.0cm}{\footnotesize{\bf Fig~2.}  
The coordinate system. The design of this figure is taken from Ref. \cite{9}.}
\label{fig2}
\end{center}

\par
Two images of the accretion disk are derived on the observation plane, i.e., the direct image with coordinates $(b^{(d)},\alpha)$ and the secondary image with coordinates $(b^{(s)},\alpha+\pi)$, as reported in \cite{9}. Assuming that the deflection angle from $M$ to the observer is $\gamma$ and the observer's inclination angle is $\theta_{0}$, we can derive the following equation:
\begin{equation}
\label{3-1}
\cos \alpha = \cos \gamma \sqrt{\cos^{2} \alpha + \cot^{2} \theta_{0}}.
\end{equation}
For the direct image of the accretion disk, Eq. (\ref{2-11}) can be rewritten as
\begin{equation}
\label{3-2}
\gamma = \frac{1}{\sqrt{2M}} \int_{0}^{1/r}\frac{1}{\sqrt{G(u)}}{\rm d}u = 2 \sqrt{\frac{P}{Q}} \Big(F(\zeta_{\rm r},k) - F(\zeta_{\rm \infty}, k)\Big),
\end{equation}
where $F(\zeta_{\rm r},k)$ and $F(\zeta_{\rm \infty})$ are the elliptical integrals, and $k^{2}=\frac{Q-P+6M}{2Q}$, $\sin^{2}\zeta_{\rm r}=\frac{Q-P+2M+4MP/r}{Q-P+6M}$, and $\sin^{2}\zeta_{\rm \infty}=\frac{Q-P+2M}{Q-P+6M}$ \cite{9}. Thus, the radius $r$ as a function of $\alpha$ and $P$ is obtained; we have
\begin{equation}
\label{3-3}
\frac{1}{r} = \frac{P-2M-Q}{4MP} + \frac{Q-P+6M}{4MP} sn^{2}\Bigg(\frac{\gamma}{2}\sqrt{\frac{Q}{P}} + F(\zeta_{\rm \infty}, k)\Bigg).
\end{equation}
According to this equation, the iso-radial curves for a given angle $\theta_{0}$ are presented. For the $(n+1)th$ order image of the accretion disk, Eq. (\ref{3-2}) satisfies
\begin{equation}
\label{3-4}
2n \pi - \gamma = 2 \sqrt{\frac{P}{Q}} \Big(2K(k)- F(\zeta_{\rm r},k) - F(\zeta_{\rm \infty}, k)\Big),
\end{equation}
where $K(k)$ is the complete elliptic integral.
\begin{center}
\includegraphics[width=5cm,height=5cm]{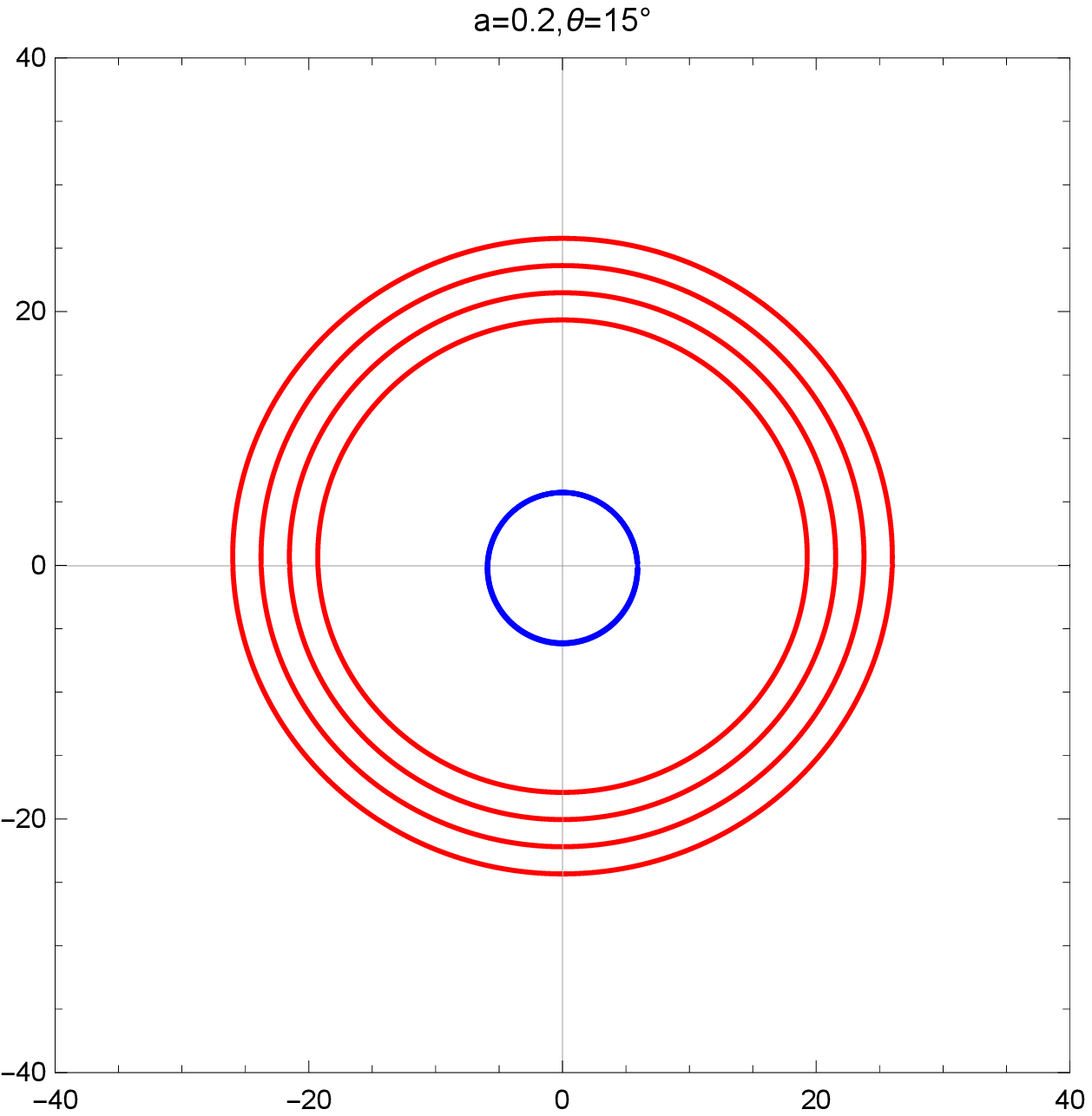}
\includegraphics[width=5cm,height=5cm]{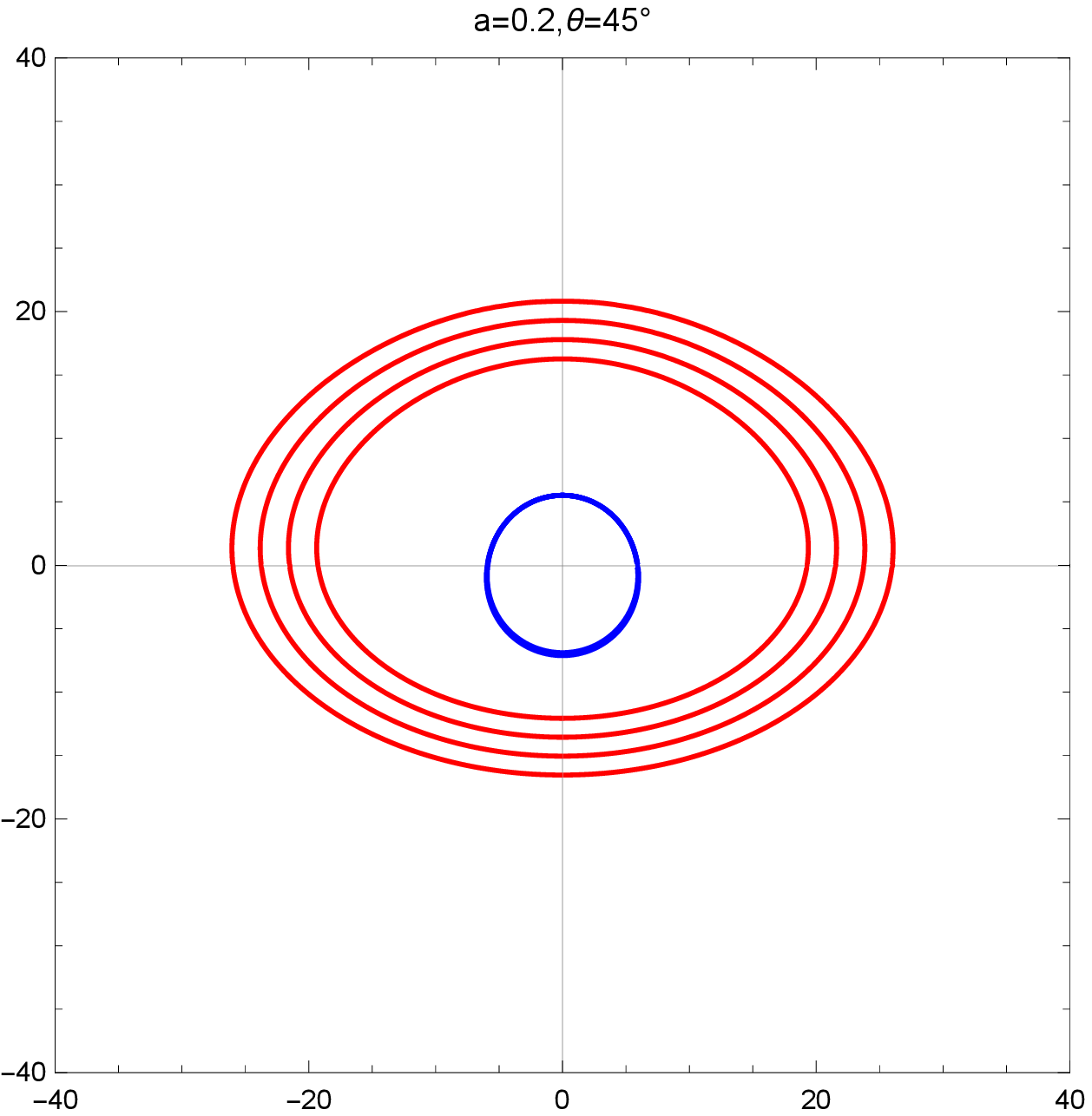}
\includegraphics[width=5cm,height=5cm]{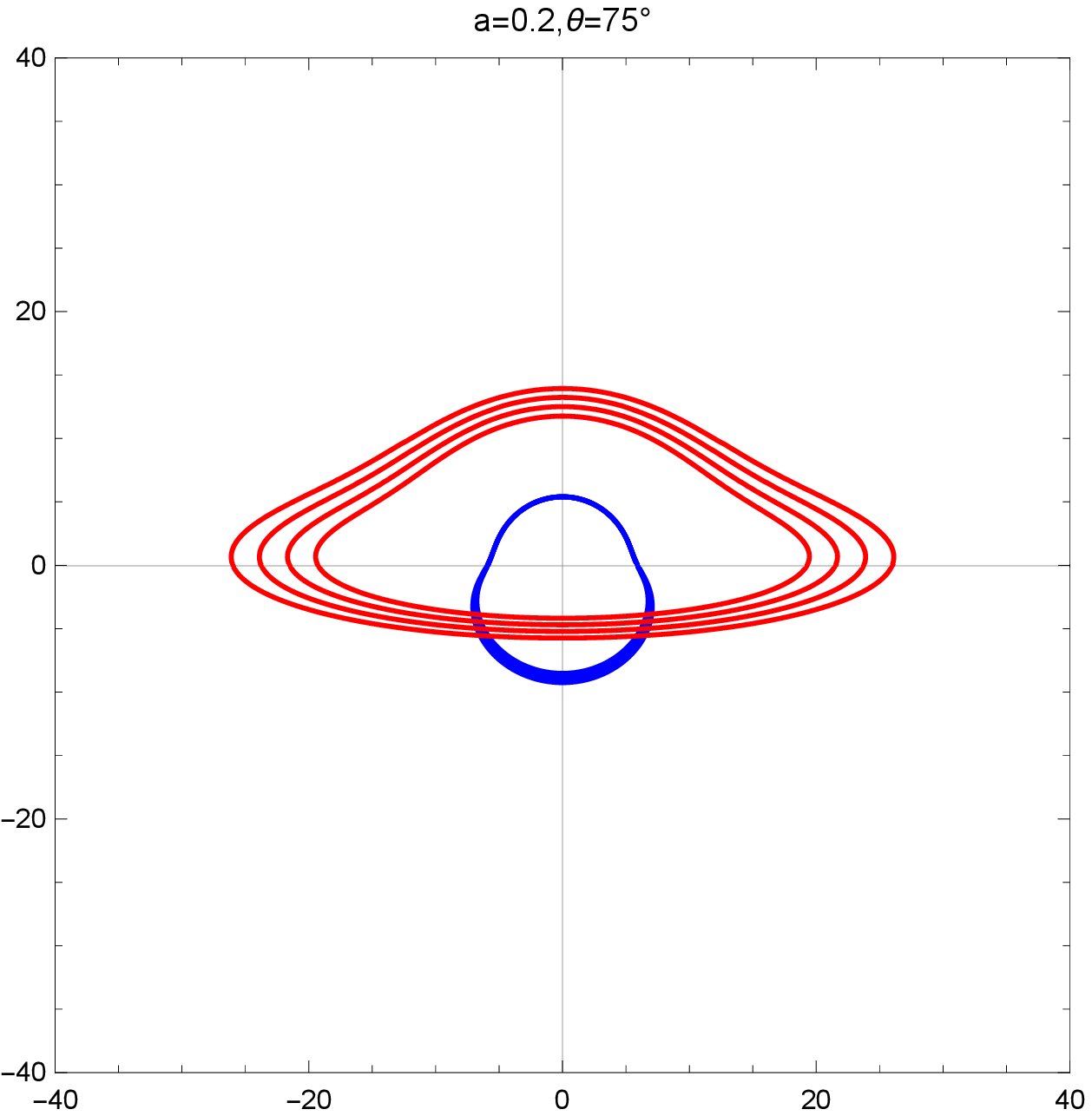}
\includegraphics[width=5cm,height=5cm]{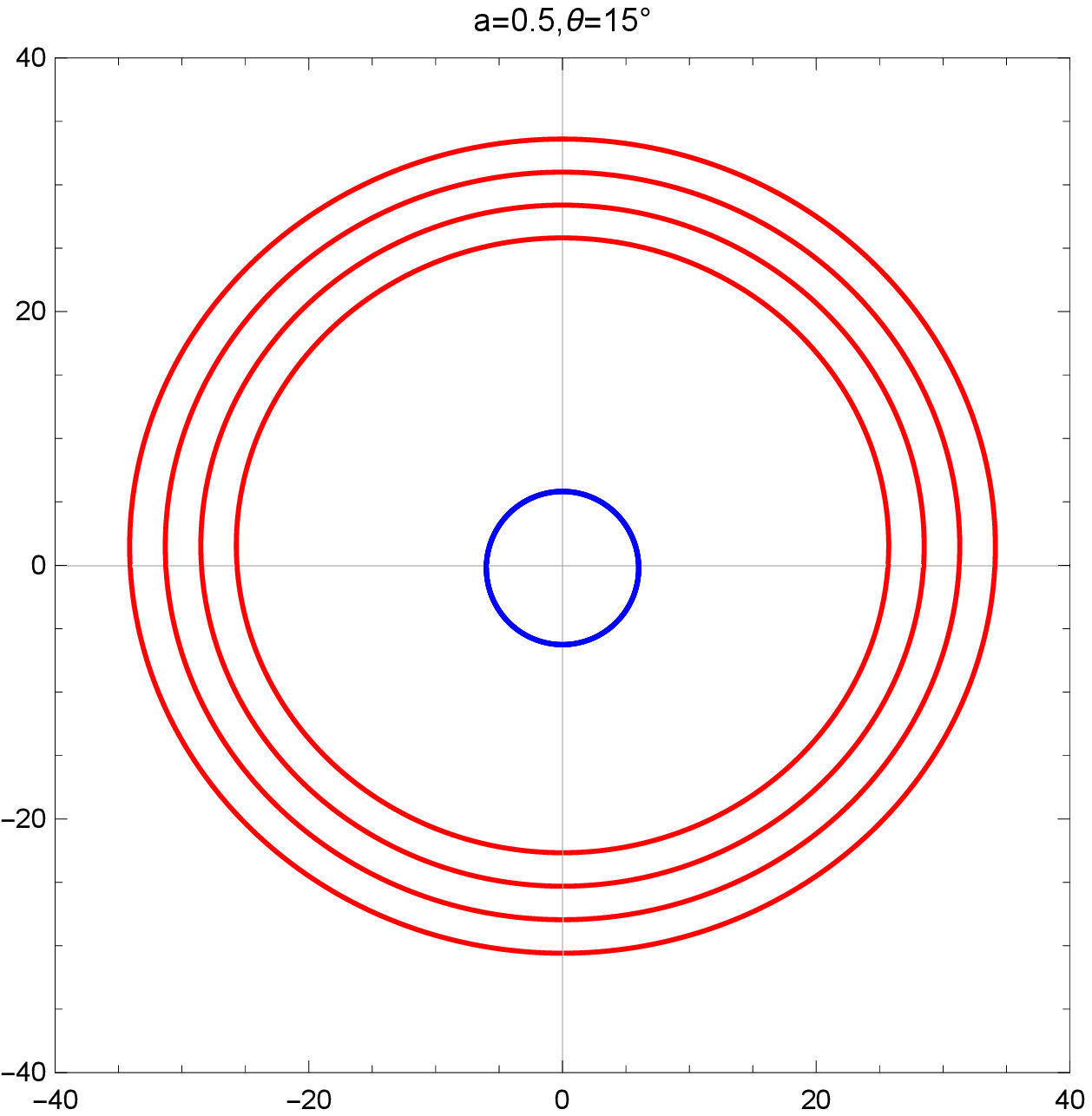}
\includegraphics[width=5cm,height=5cm]{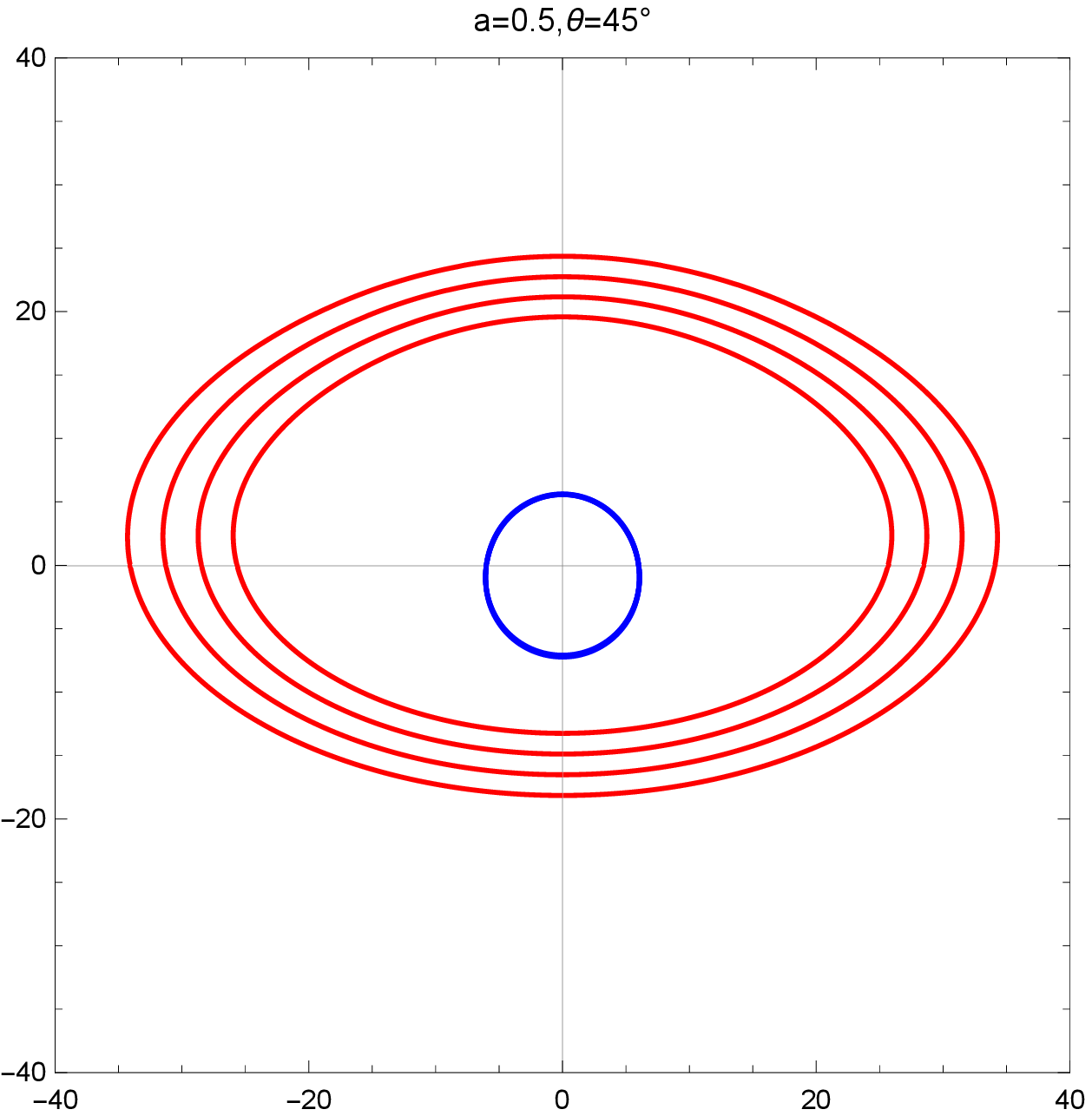}
\includegraphics[width=5cm,height=5cm]{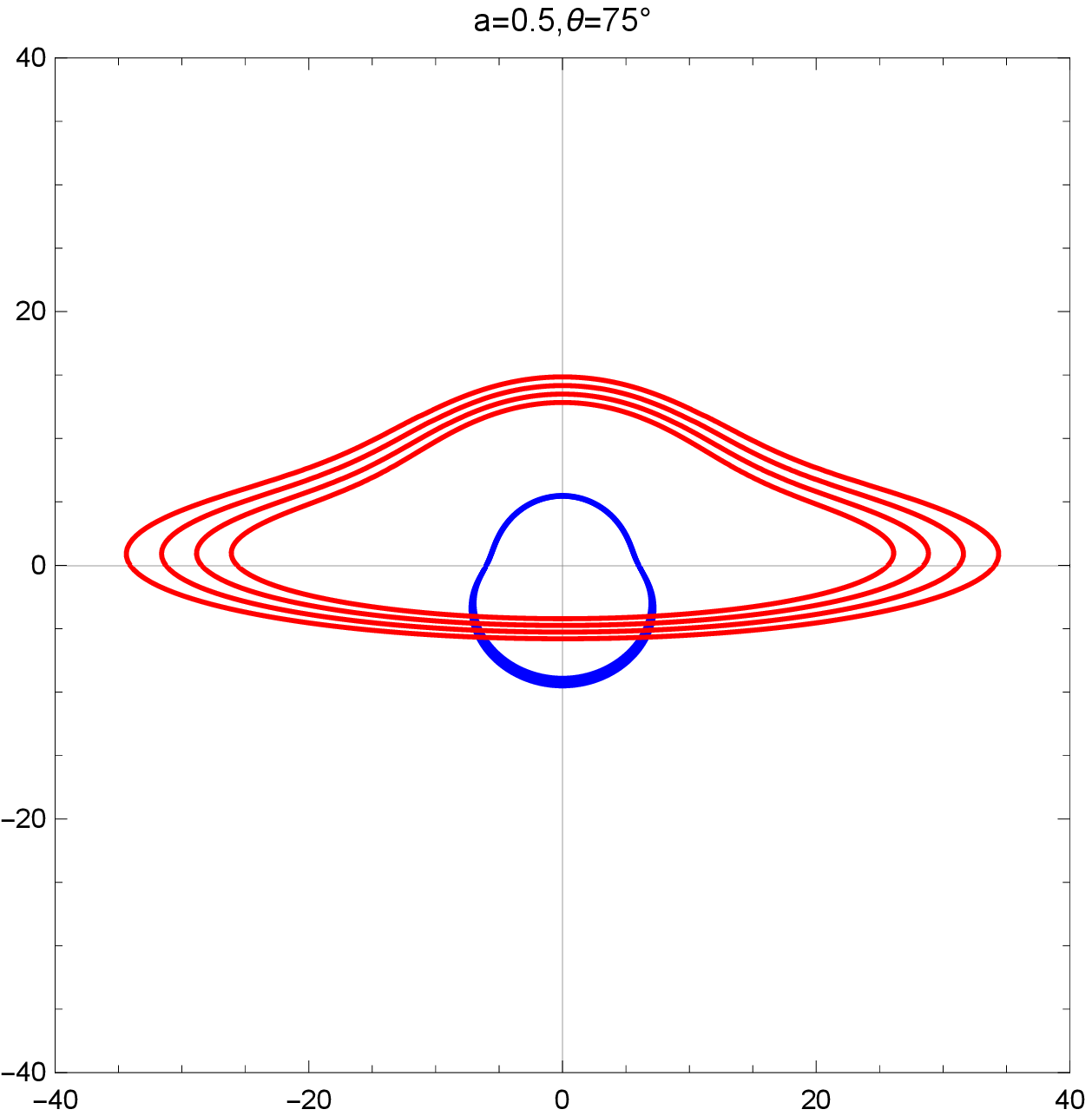}
\parbox[c]{15.0cm}{\footnotesize{\bf Fig~3.}  
Direct and secondary images of the thin accretion disk around the Schwarzschild string cloud BH with the following inclination angles of the observer: $\theta_{0}=15^{\circ},45^{\circ},75^{\circ}$. Red curves represent direct image, and blue curves represent the secondary image of the disk. {\em Top Panel} -- string cloud parameter $a=0.2$ and {\em Bottom Panel} -- string cloud parameter $a=0.5$. We set $M=1$.}
\label{fig3}
\end{center}

\par
Figure 3 displays the direct and secondary images of circular rings orbiting a Schwarzschild string cloud BH, located at various distances from the center ($r=16M$, $r=18M$, $r=20M$, and $r=22M$, from inside to outside) and at different inclination angles ($\theta_{0}=15^{\circ}$, $45^{\circ}$, and $75^{\circ}$). The direct images correspond to photons emitted in directions above the equatorial plane, and the secondary images correspond to photons emitted in directions below the equatorial plane. It is evident that as the string cloud parameter ($a$) increases, the direct image expands outward, increasing in the shadow radius. Conversely, the shape and size of the secondary images decrease. Additionally, an increased inclination angle results in a clearer separation between the direct and secondary images.

\section{Optical appearance of Schwarzschild string cloud BH}
\label{sec:4}
\par
According to Ref. \cite{57}, the radial dependence of energy flux radiated by a thin accretion disk around a BH is derived. The flux of the radiant energy is given by \cite{58}
\begin{equation}
\label{4-1}
F= - \frac{\dot{M}}{4\pi \sqrt{\rm -g}} \frac{\Omega_{,\rm r}}{(E-\Omega L)^{2}} \int_{r_{\rm in}}^{r} (E- \Omega L)L_{,\rm r} {\rm d} r,
\end{equation}
where $\dot{M}$ represents the mass accretion rate, $\rm g$ represents the determinant of the metric, and $r_{\rm in}$ represents the inner edge of the accretion disk. $E$, $\Omega$, and $L$ represent the energy, angular momentum, and angular velocity, respectively.

\par
For a static spherically symmetric metric ${\rm d}s^{2}=g_{\rm tt} {\rm d}t^{2} + g_{\rm \phi\phi} {\rm d}\phi^{2} + g_{\rm rr}{\rm d}r^{2} + g_{\rm \theta\theta}{\rm d}\theta^{2}$, $E$, $L$, and $\Omega$ are expressed as follows:
\begin{equation}
\label{4-2}
E=-\frac{g_{\rm tt}}{\sqrt{-g_{\rm tt}-g_{\rm \phi\phi}\Omega^{2}}},~~L=\frac{g_{\rm \phi\phi}\Omega}{\sqrt{-g_{\rm tt}-g_{\rm \phi\phi}\Omega^{2}}},
\end{equation}
\begin{equation}
\label{4-3}
\Omega=\frac{{\rm d}\phi}{{\rm d}t}=\sqrt{-\frac{g_{\rm tt,r}}{g_{\rm \phi\phi,r}}}.
\end{equation}
In our situation, the flux of the radiant energy over the disk is given as
\begin{equation}
\label{4-4}
F = \frac{3 \dot{M} \sqrt{\frac{M}{r^{3}}}}{r (24 M \pi - 8 \pi r + 8 a \pi r)} \int_{r_{\rm in}}^{r} \frac{\sqrt{\frac{M}{r^{3}}} r \Big(6M + a r - r \Big)}{6 M + 2 r \Big(a - 1 \Big)}{\rm d}r.
\end{equation}

\par
The observed flux $F_{\rm obs}$ differs from the source flux $F$ owing to the effect of redshift. As mentioned in Ref. \cite{9}, the observed flux $F_{\rm obs}$ can be expressed as follows:
\begin{equation}
\label{4-5}
F_{\rm obs} = \frac{F}{(1+z)^{4}}.
\end{equation}
When a photon is emitted by a particle orbiting around a BH, the energy of the photon can be obtained by projecting its 4-momentum, which is denoted as $p$, onto the 4-velocity, which is denoted as $\mu$, of the emitting particle \cite{13}. Therefore, we have
\begin{equation}
\label{4-6}
E_{\rm em} = p_{\rm t} \mu^{\rm t} + p_{\rm \phi} \mu^{\rm \phi} = p_{\rm t} \mu^{\rm t} \Bigg(1 + \Omega \frac{p_{\rm \phi}}{p_{\rm t}}\Bigg),
\end{equation}
where $p_{\rm t}$ and $p_{\rm \phi}$ represent the energy ($p_{\rm t}=-E$) and angular momentum ($p_{\rm \phi}=L$) of the photon, respectively, which are conserved along the geodesic owing to the Killing symmetries of the spacetime. When the observer is located at infinity, the ratio of $p_{\rm t}$ to $p_{\rm \phi}$ gives the impact parameter of the photon relative to the z-axis. Using Eq. (\ref{3-1}), we have
\begin{equation}
\label{4-7}
\sin \theta_{0} \cos \alpha = \cos \gamma  \sin \beta,
\end{equation}
and one can obtain
\begin{equation}
\label{4-8}
\frac{p_{\rm t}}{p_{\rm \phi}} = b \sin \theta_{0} \sin \alpha.
\end{equation}
Hence, the redshift factor is
\begin{equation}
\label{4-9}
1 + z = \frac{E_{\rm em}}{E_{\rm obs}} = \frac{1 + b \Omega \cos \beta}{\sqrt{-g_{\rm tt} - 2 g_{\rm t \phi} - g_{\rm \phi\phi}}}.
\end{equation}
For the Schwarzschild BH within the string cloud context, the above equation can be written as
\begin{equation}
\label{4-10}
1+z = \frac{\sqrt{1 - a - \frac{3M}{r}} r \big(1 + b \sqrt{\frac{M}{r^{3}}} \sin \alpha \sin \theta_{0}\big)}{3M + (a-1) r } + 1.
\end{equation}
\begin{center}
\includegraphics[width=7cm,height=5.8cm]{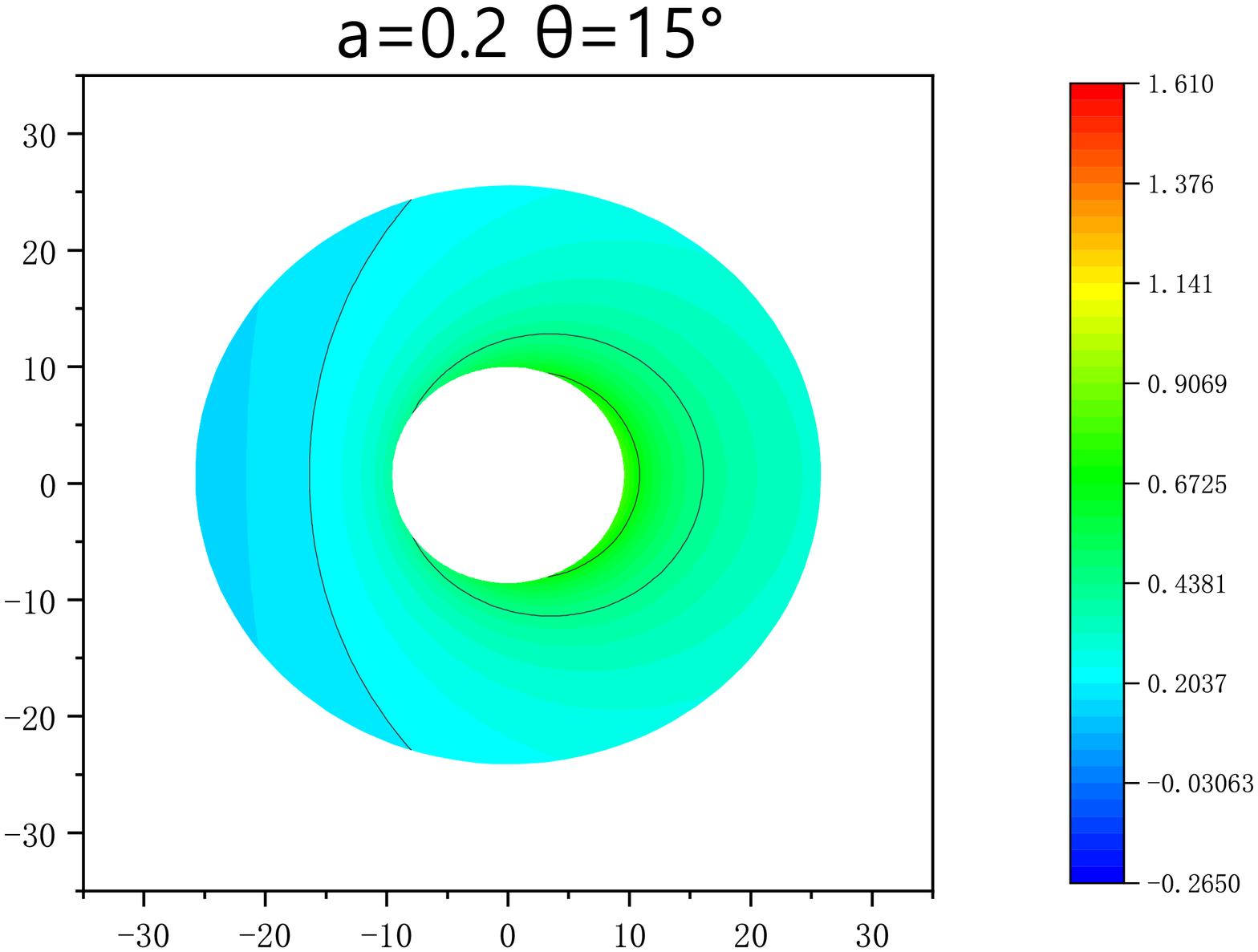}
\includegraphics[width=7cm,height=5.8cm]{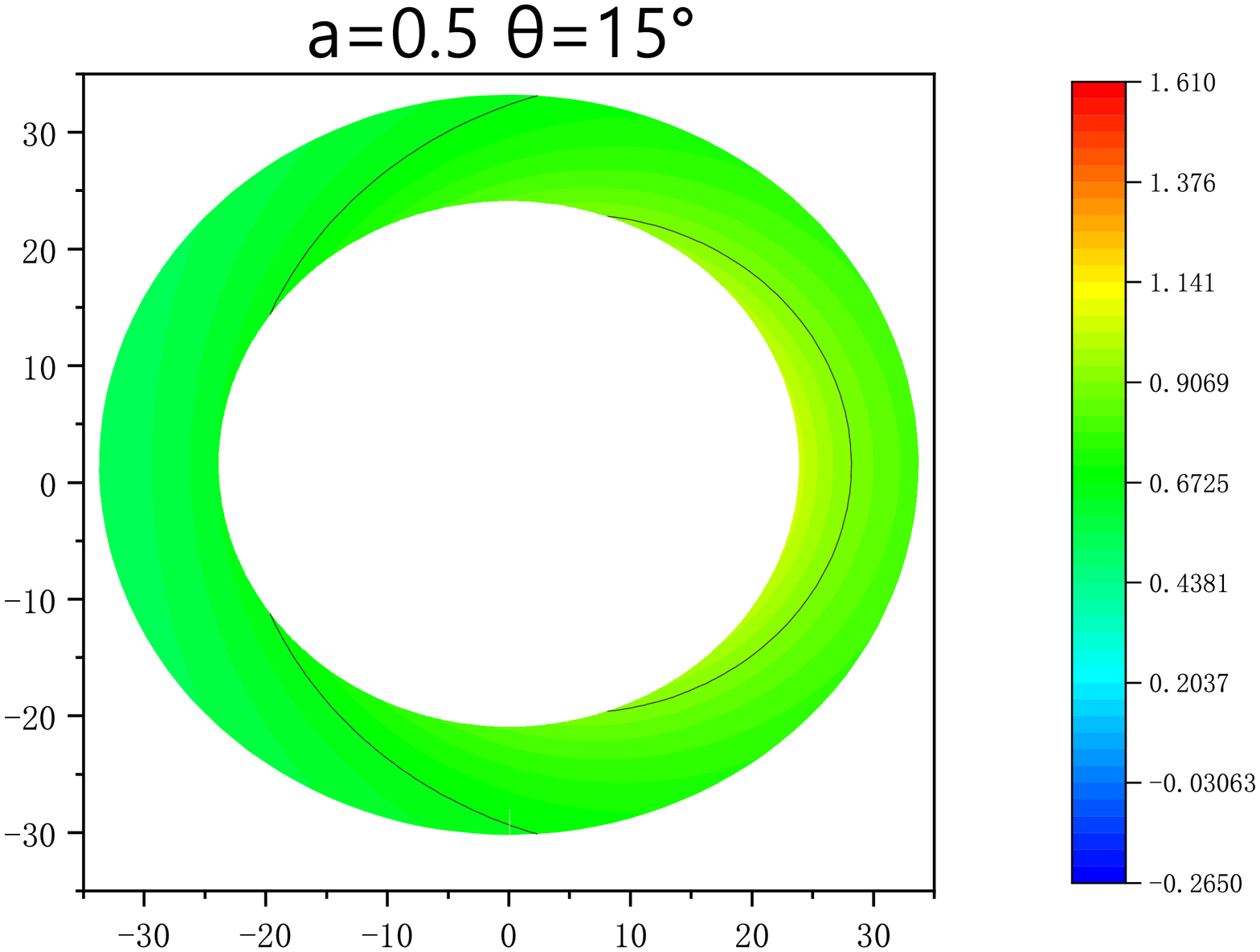}
\includegraphics[width=7cm,height=5.8cm]{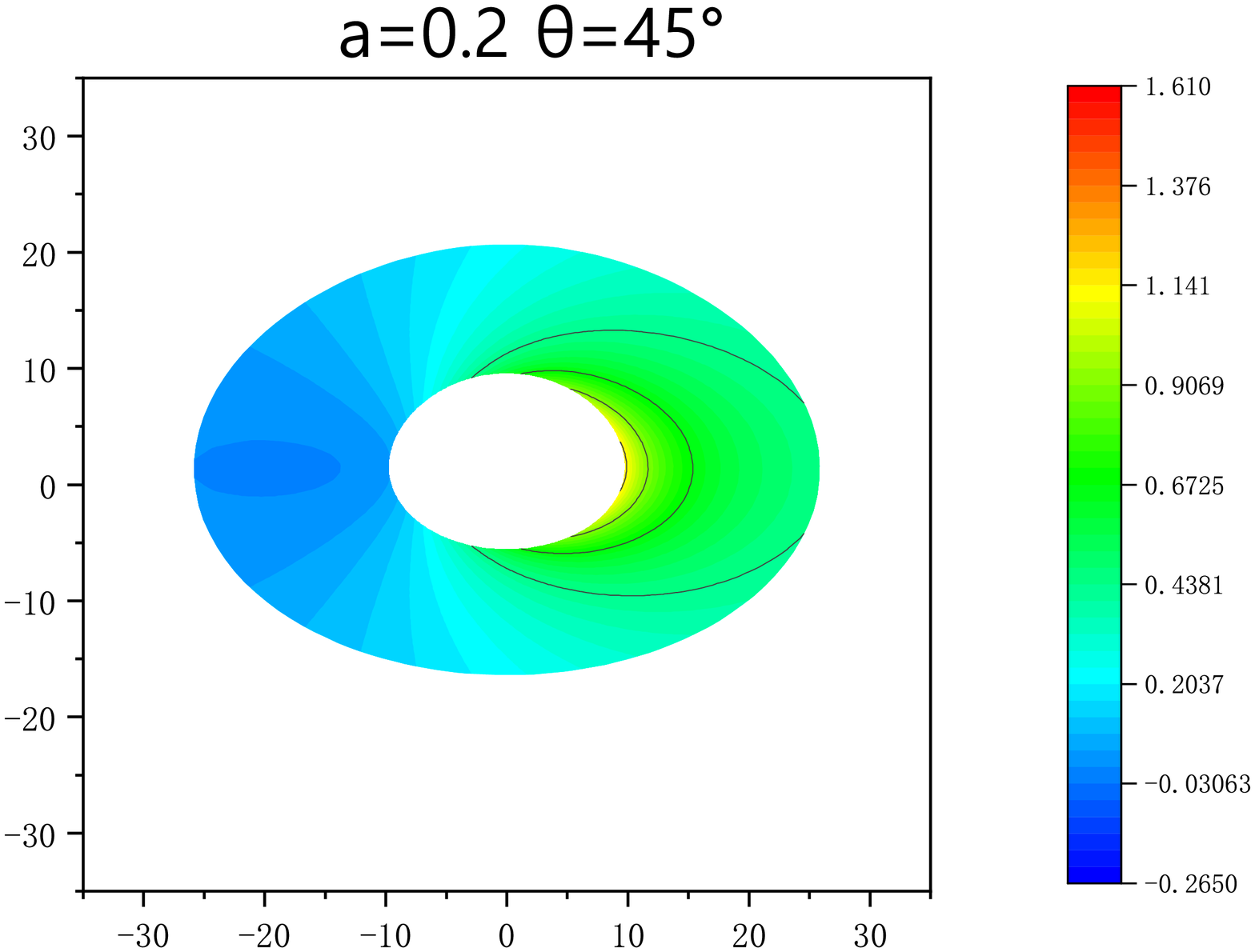}
\includegraphics[width=7cm,height=5.8cm]{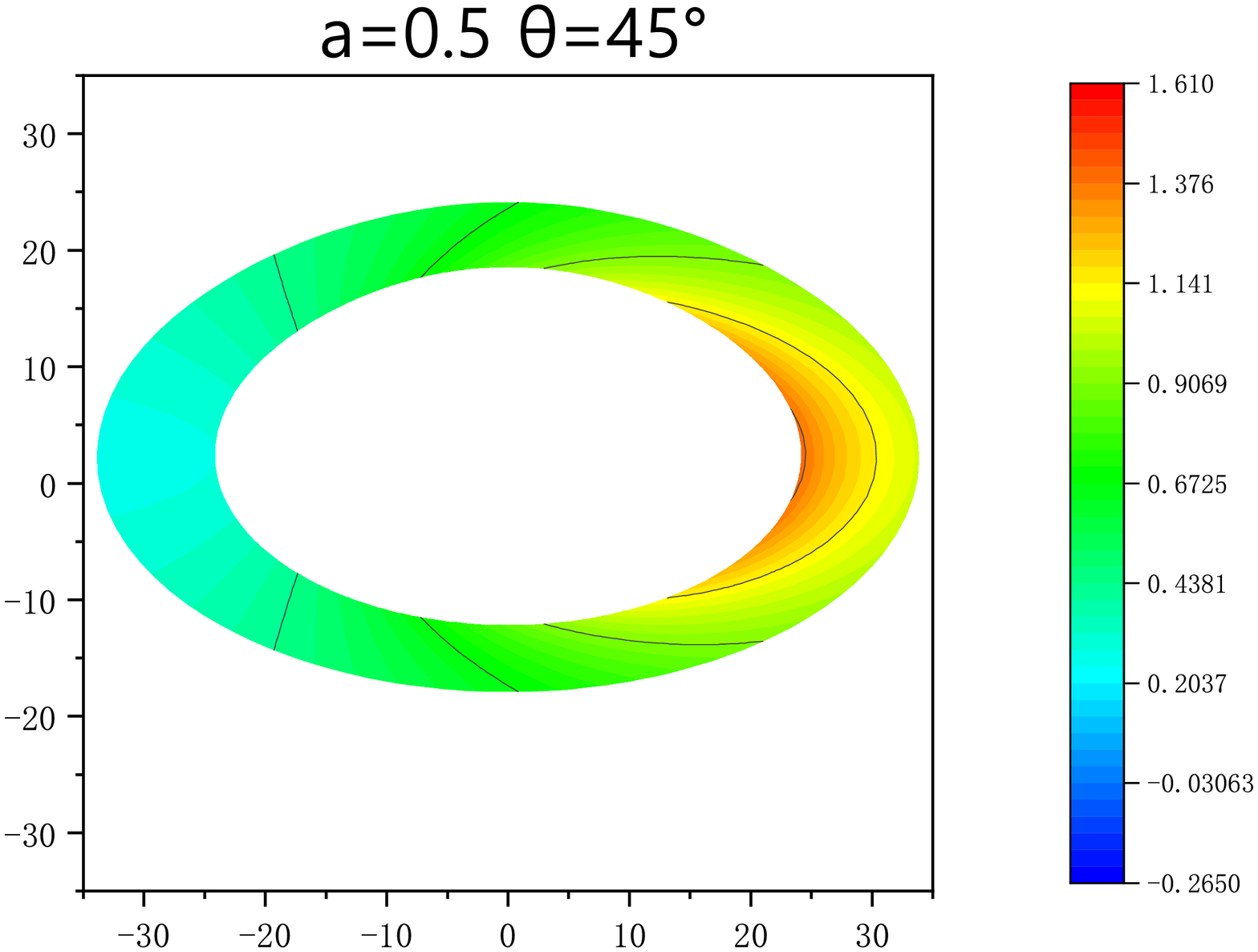}
\includegraphics[width=7cm,height=5.8cm]{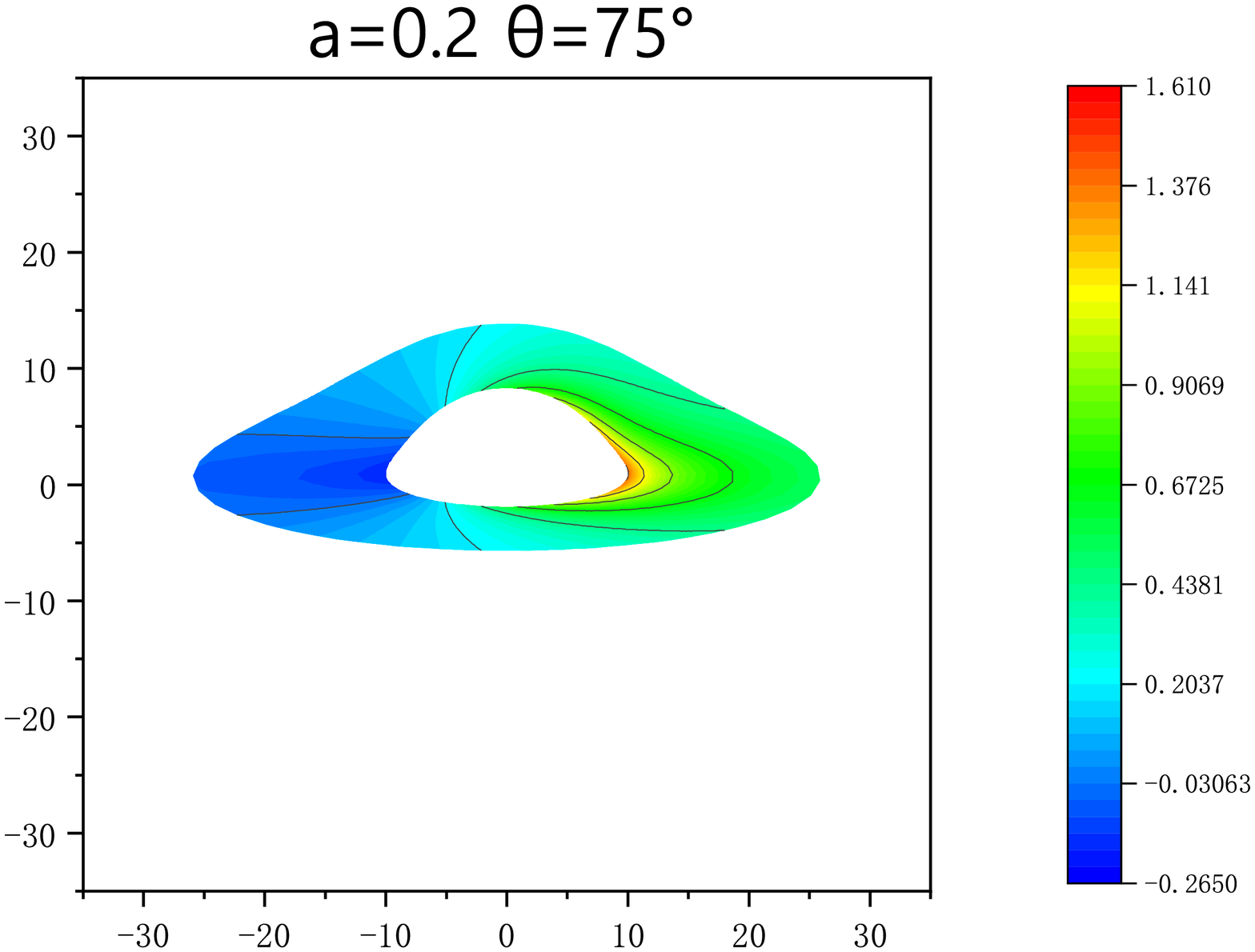}
\includegraphics[width=7cm,height=5.8cm]{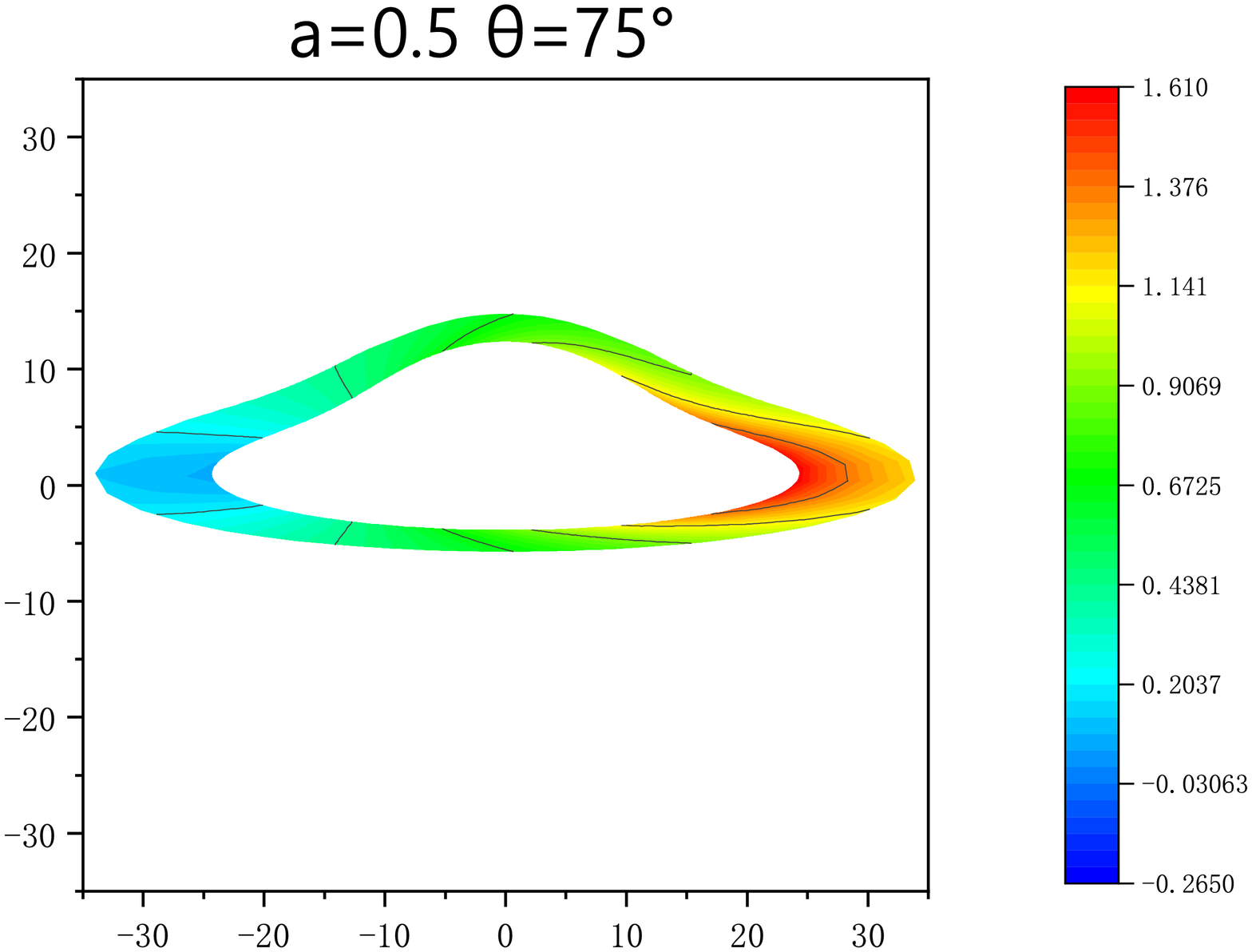}
\parbox[c]{15.0cm}{\footnotesize{\bf Fig~4.}  
Redshift distribution (curves of constant redshift z) of the thin accretion disk around the Schwarzschild string cloud BH with the following inclination angles of the observer: $\theta_{0}=15^{\circ},45^{\circ},75^{\circ}$. The inner edge of the disk is at $r_{\rm in}=16M$, and the outer edge of the disk is at $r=22M$. {\em Left Panel} -- string cloud parameter $a=0.2$ and {\em Right Panel} -- string cloud parameter $a=0.5$. The BH mass is taken as $M=1$.}
\label{fig4}
\end{center}
\begin{center}
\includegraphics[width=7cm,height=5.8cm]{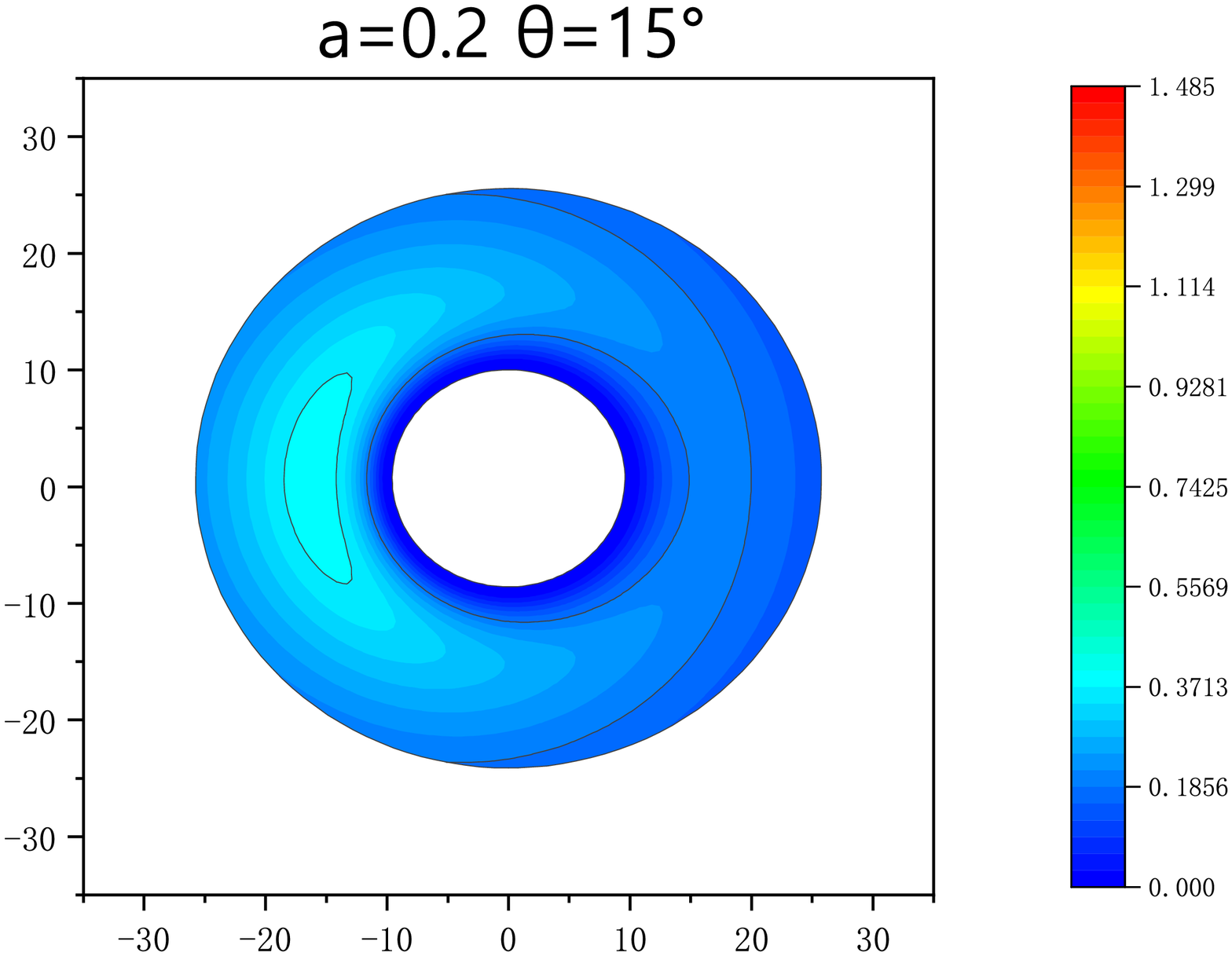}
\includegraphics[width=7cm,height=5.8cm]{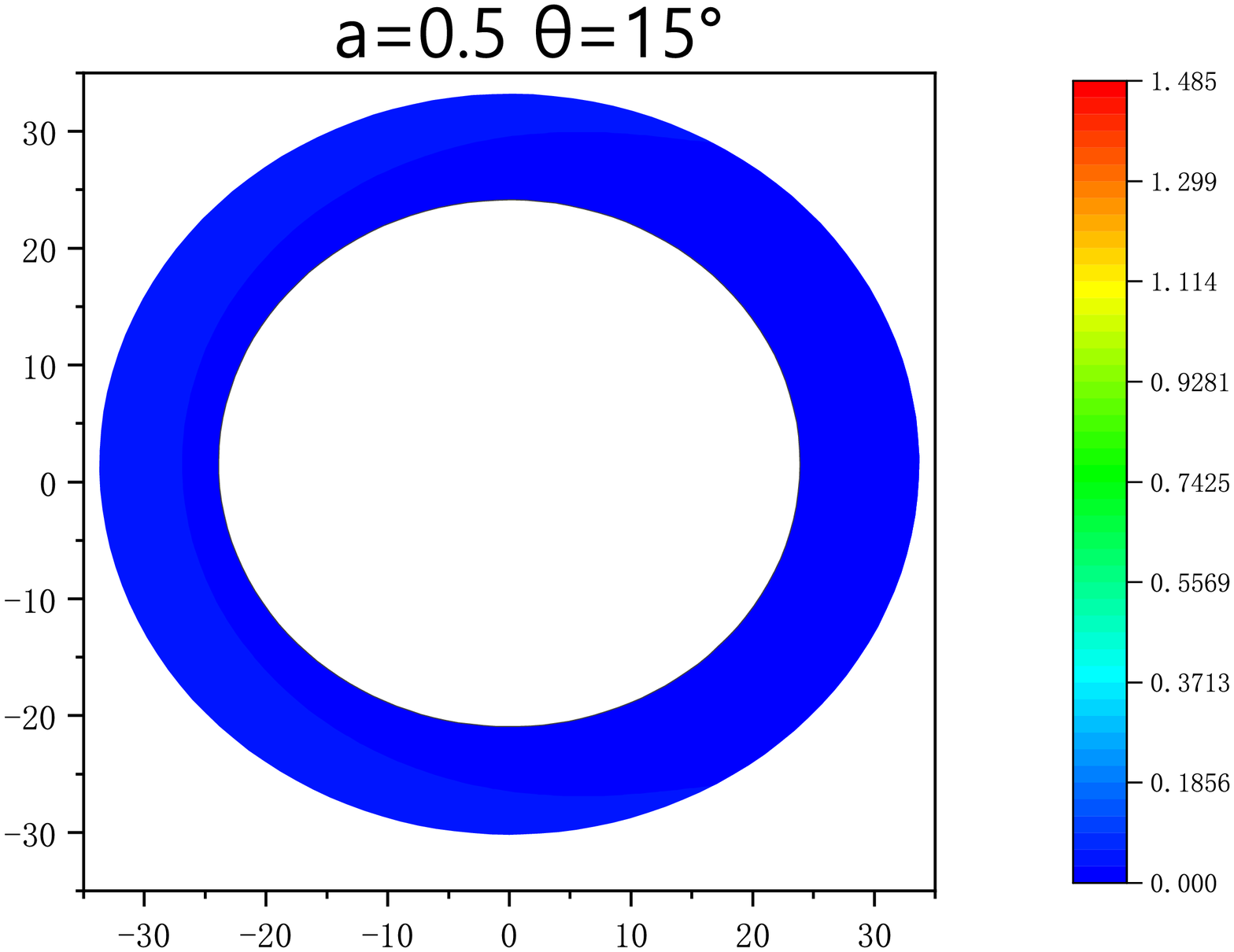}
\includegraphics[width=7cm,height=5.8cm]{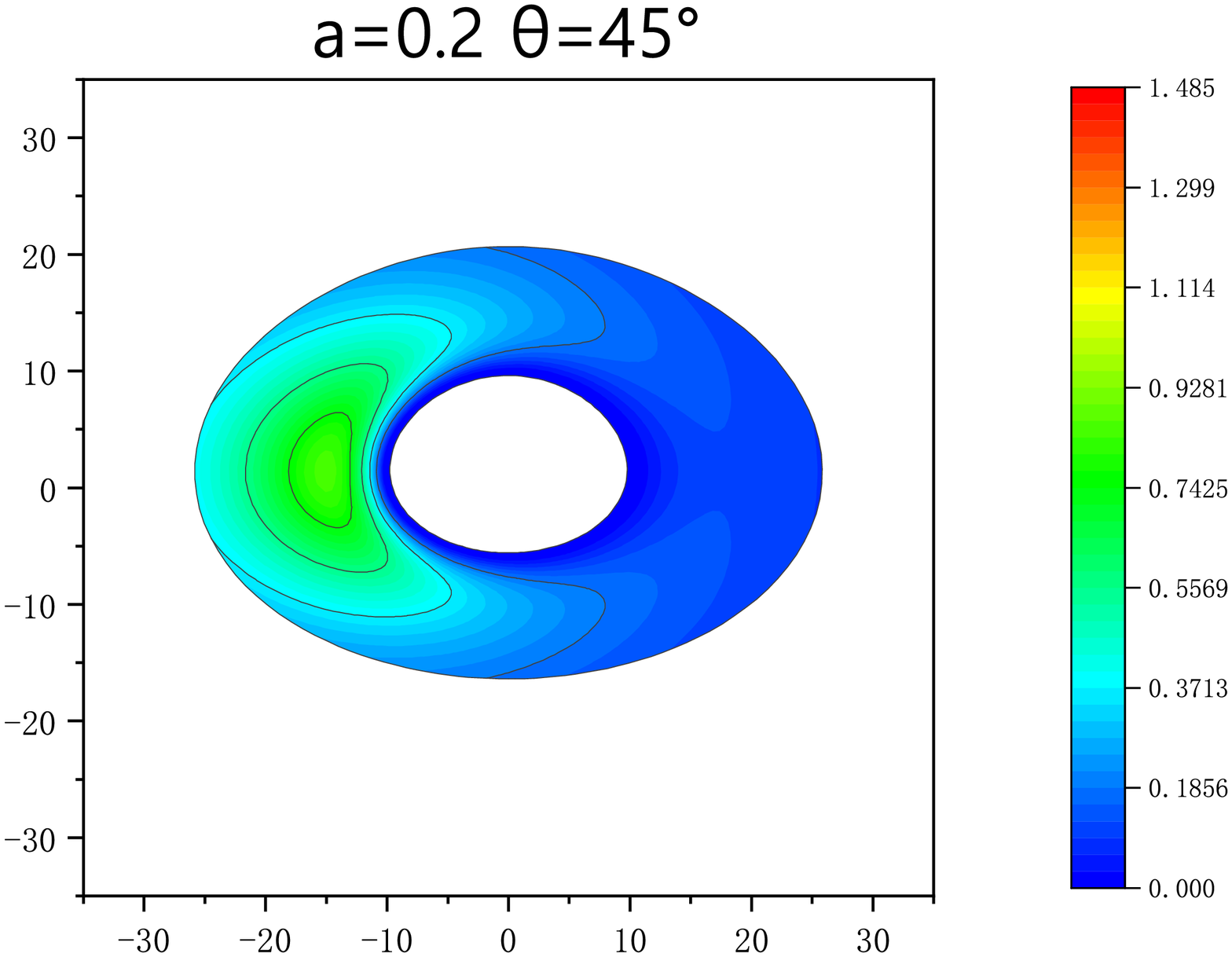}
\includegraphics[width=7cm,height=5.8cm]{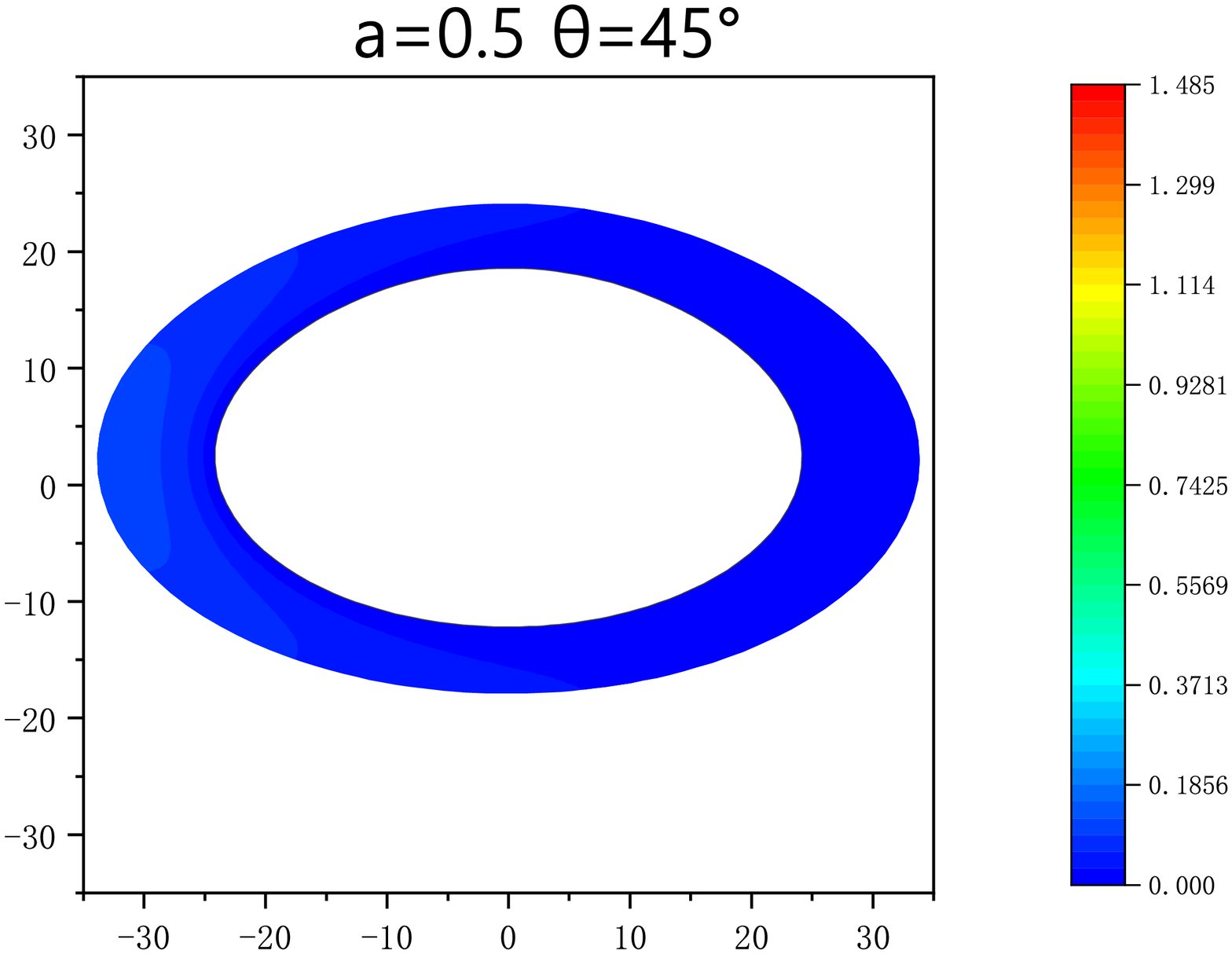}
\includegraphics[width=7cm,height=5.8cm]{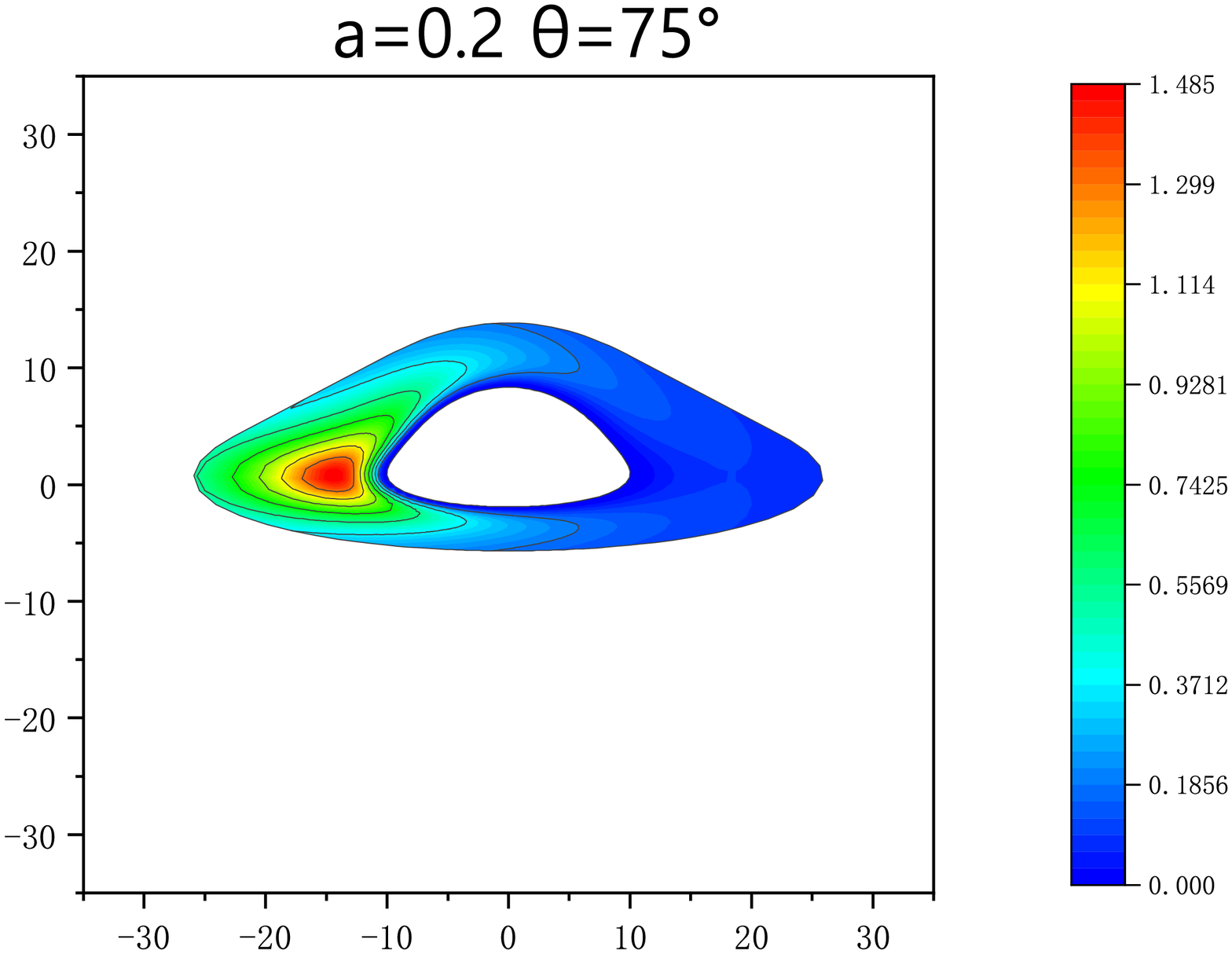}
\includegraphics[width=7cm,height=5.8cm]{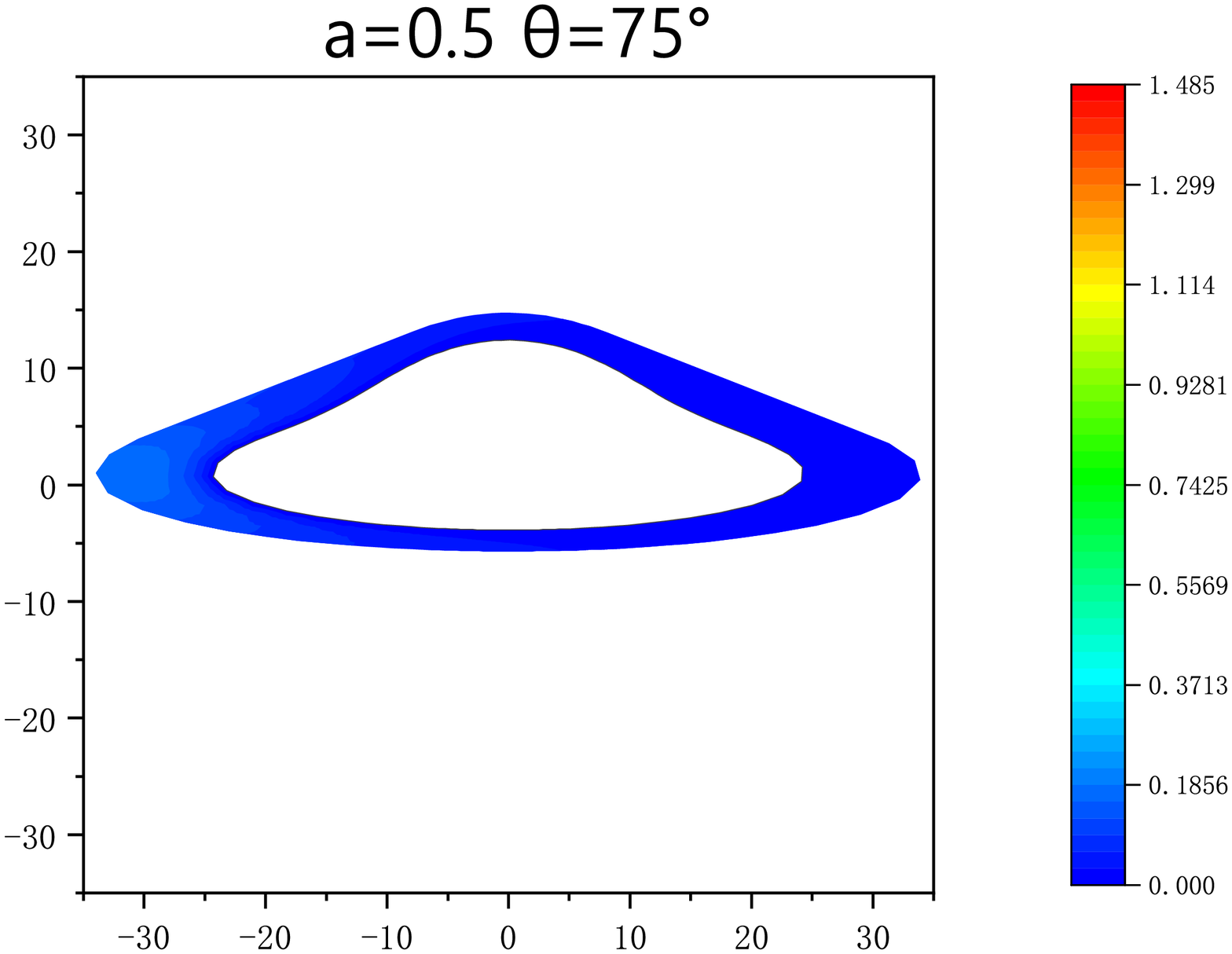}
\parbox[c]{15.0cm}{\footnotesize{\bf Fig~5.}  
Total observed flux of the direct image for the thin accretion disk around the Schwarzschild string cloud BH with the following inclination angles of the observer: $\theta_{0}=15^{\circ},45^{\circ},75^{\circ}$. The inner edge of the disk is at $r_{\rm in}=16M$, and the outer edge of the disk is at $r=22M$. {\em Left Panel} -- string cloud parameter $a=0.2$ and {\em Right Panel} -- string cloud parameter $a=0.5$. The BH mass is taken as $M=1$.}
\label{fig5}
\end{center}
\begin{center}
\includegraphics[width=7cm,height=5.8cm]{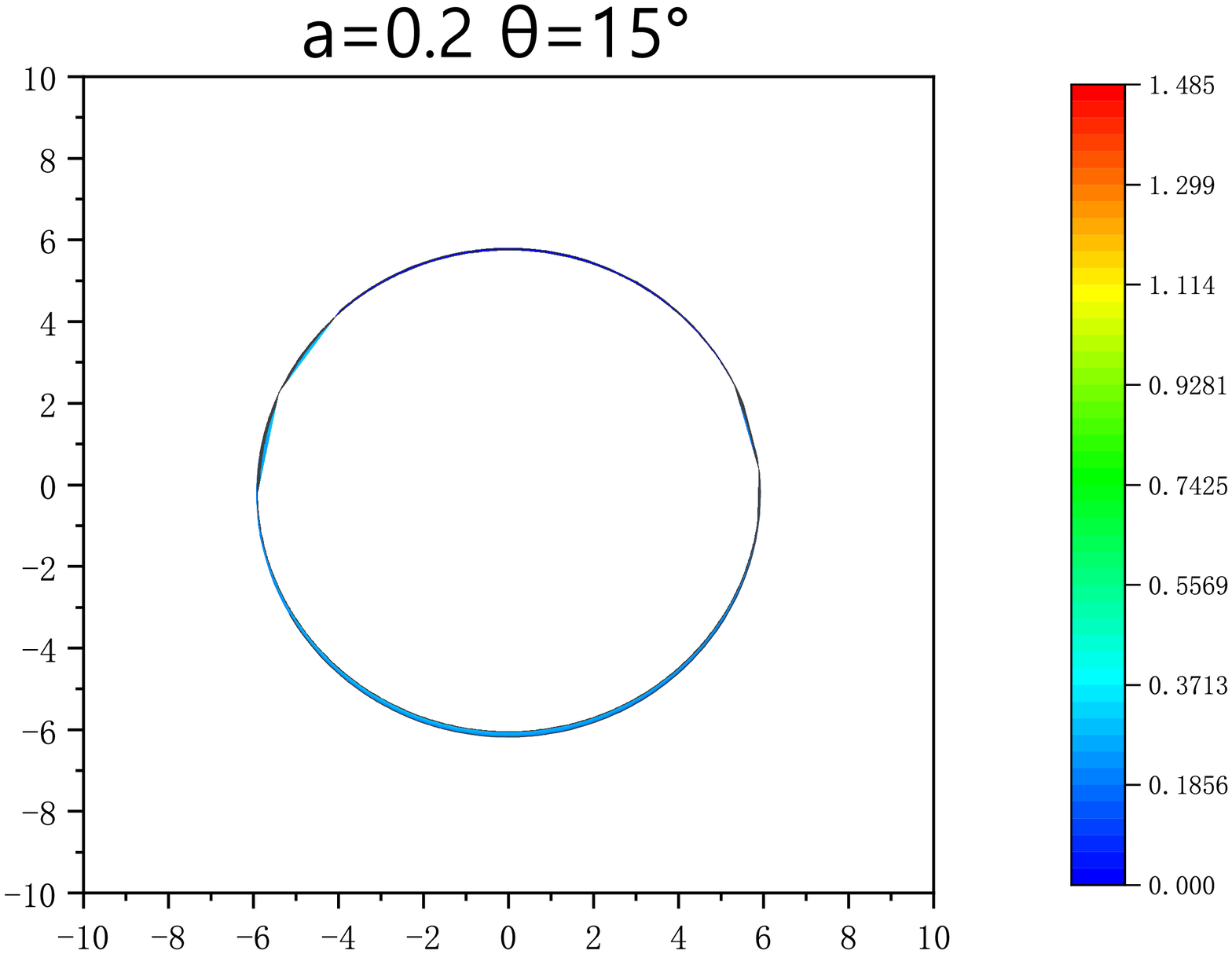}
\includegraphics[width=7cm,height=5.8cm]{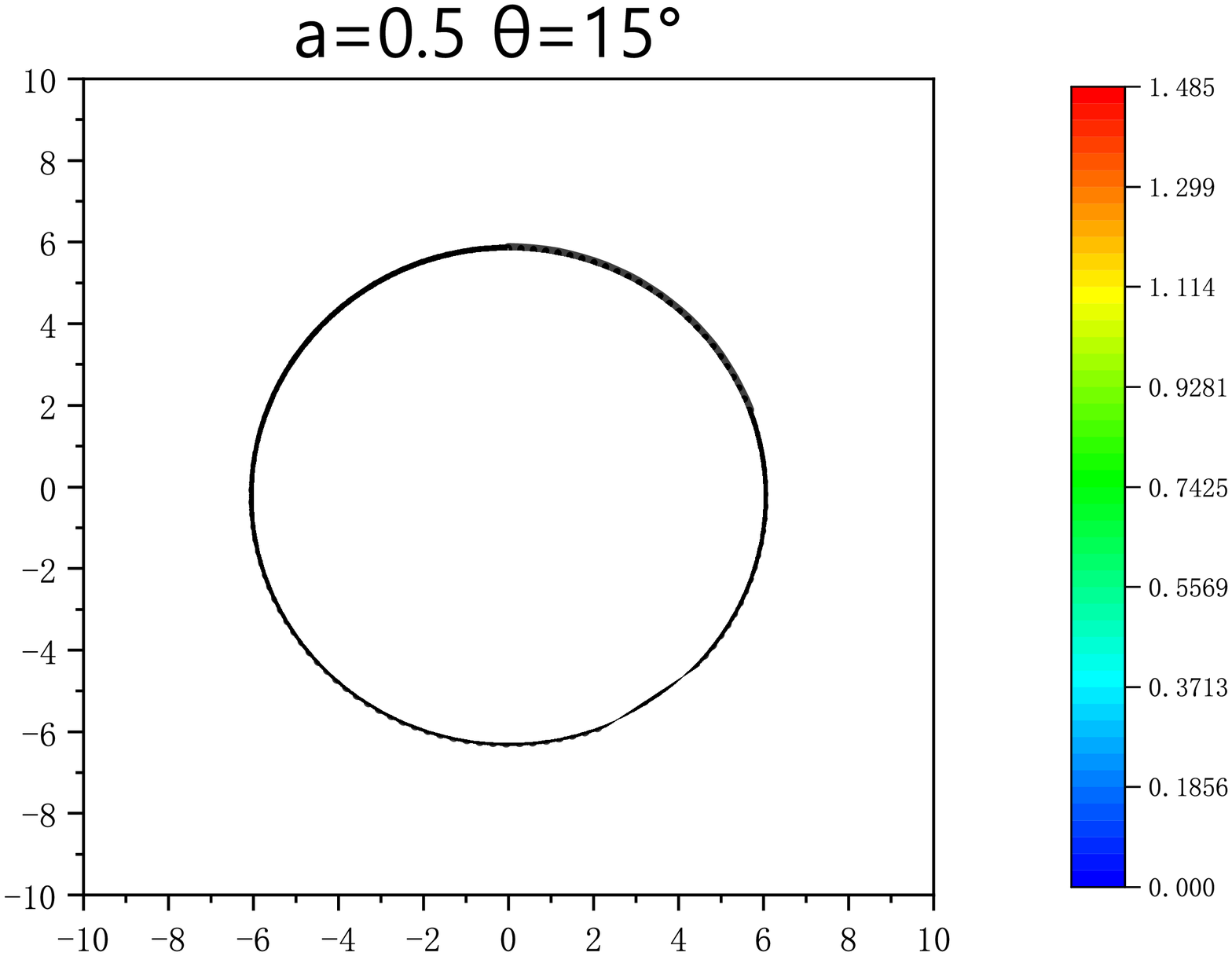}
\includegraphics[width=7cm,height=5.8cm]{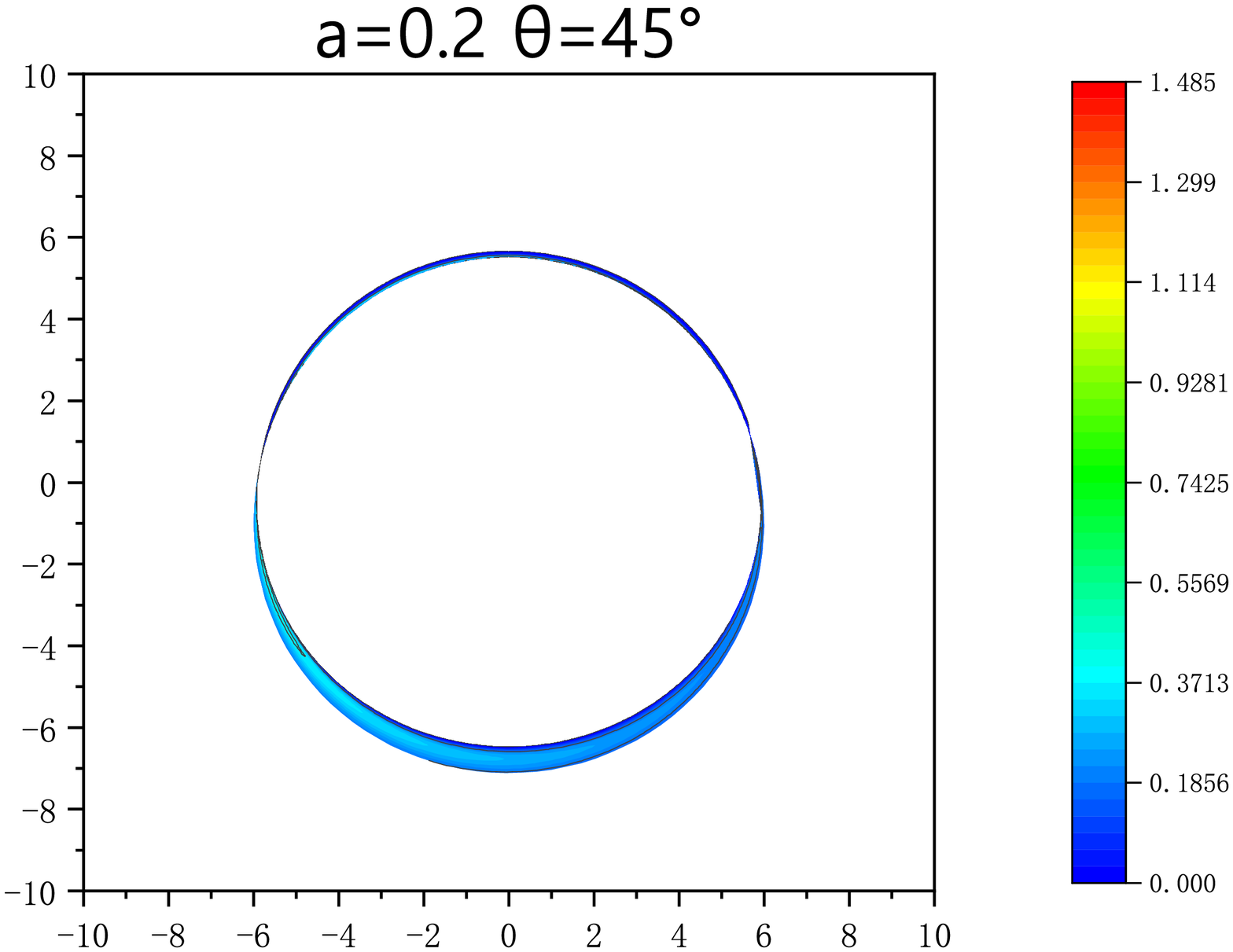}
\includegraphics[width=7cm,height=5.8cm]{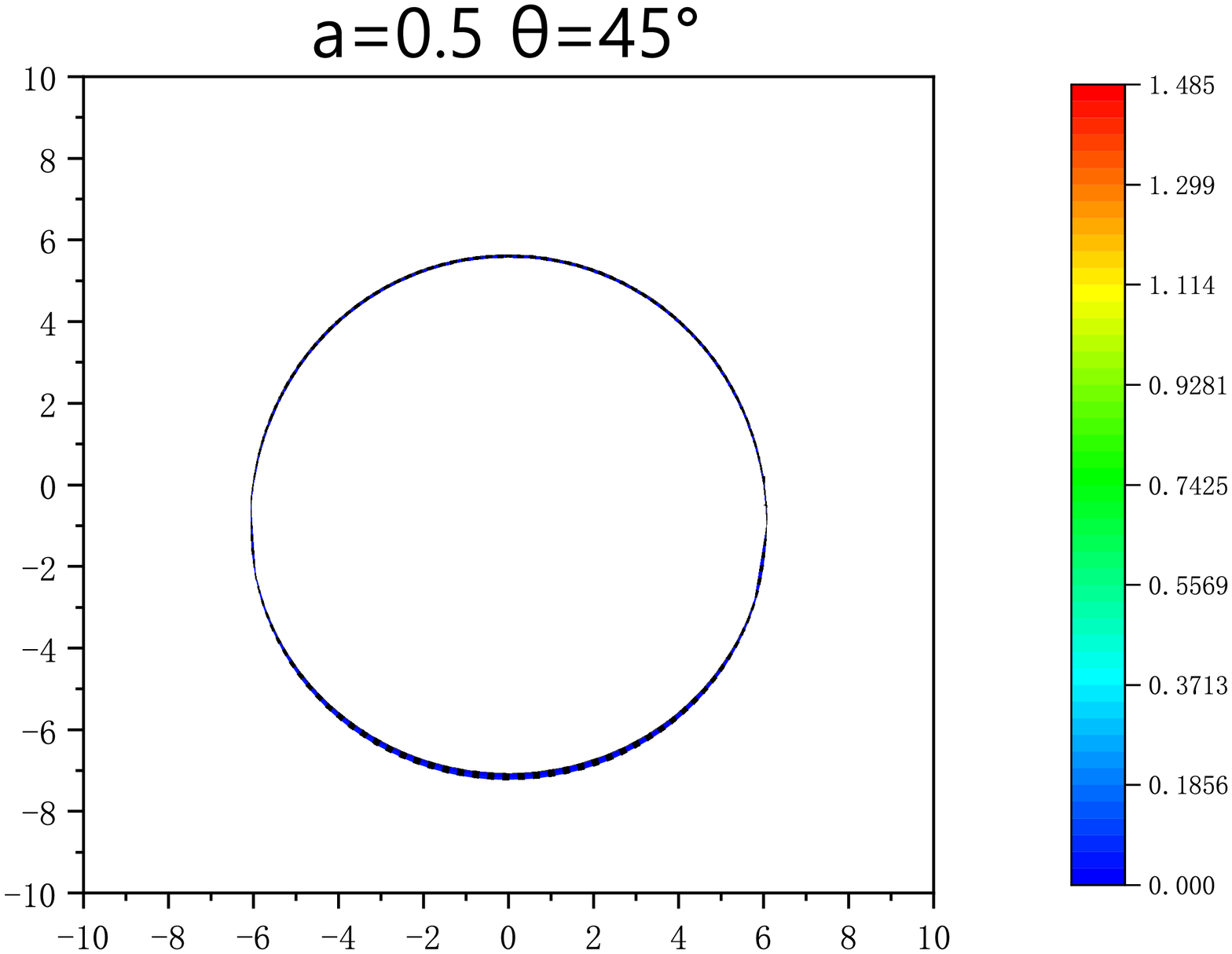}
\includegraphics[width=7cm,height=5.8cm]{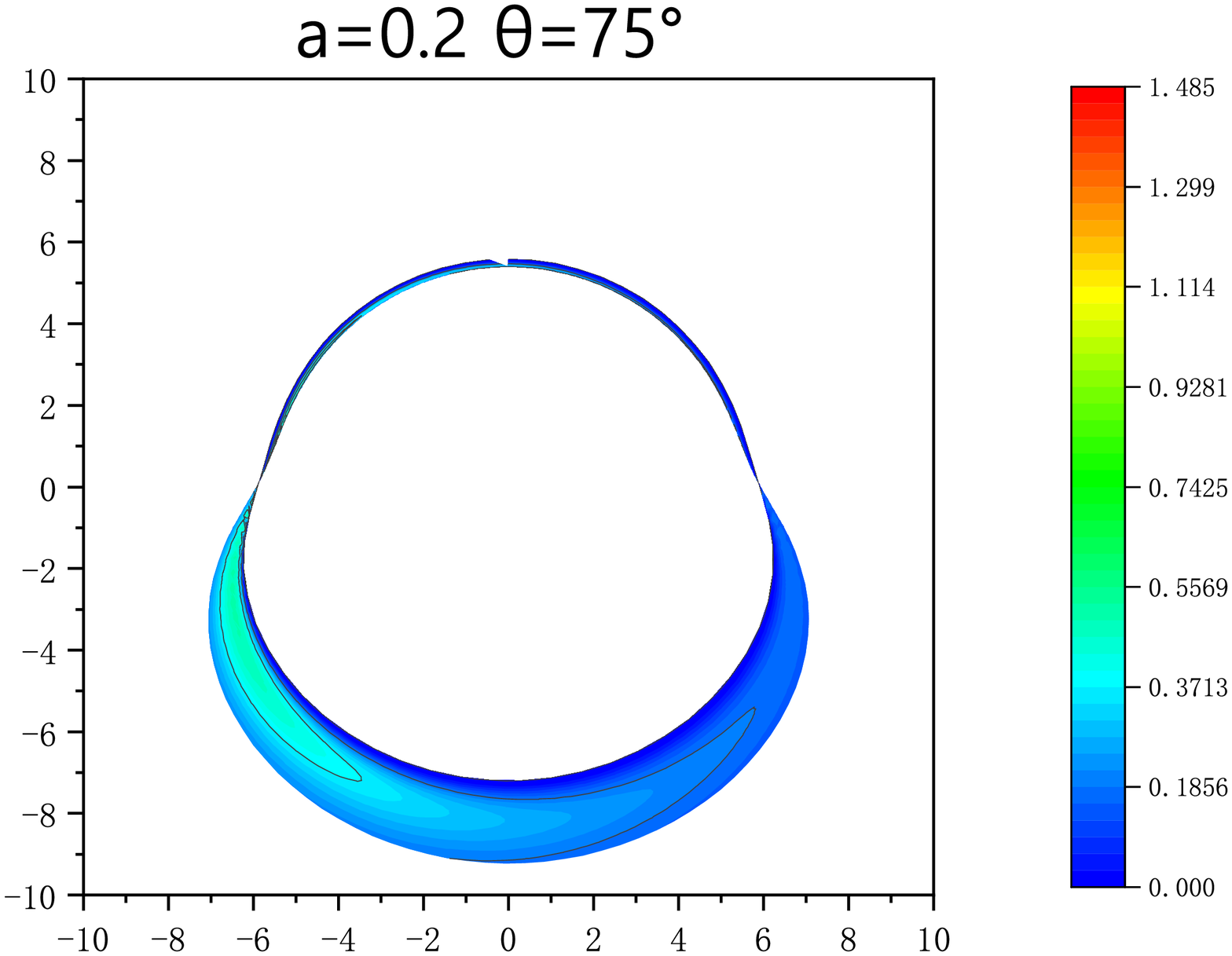}
\includegraphics[width=7cm,height=5.8cm]{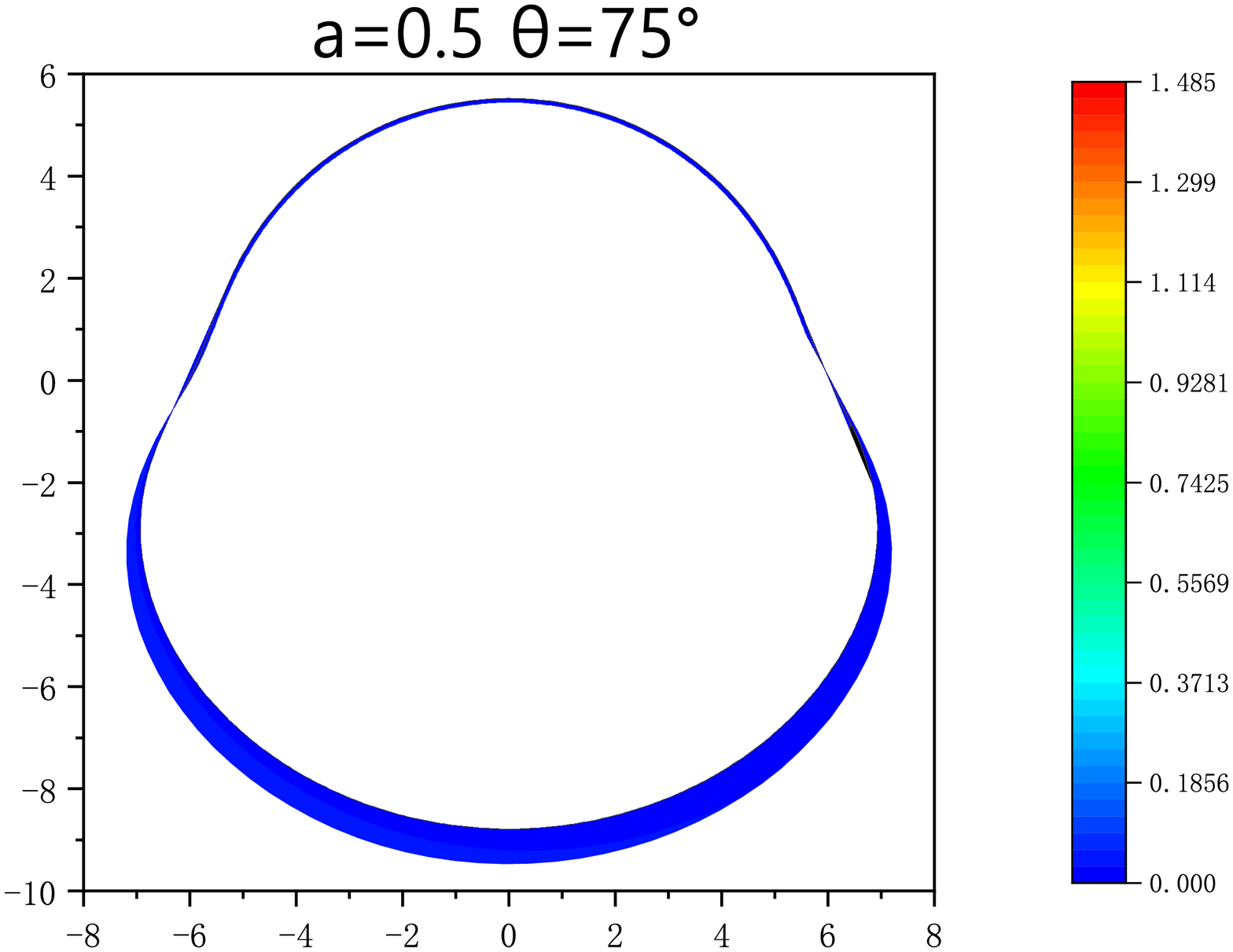}
\parbox[c]{15.0cm}{\footnotesize{\bf Fig~6.}  
Total observed flux of the secondary image for the thin accretion disk around the Schwarzschild string cloud BH with the following inclination angles of the observer: $\theta_{0}=15^{\circ},45^{\circ},75^{\circ}$. The inner edge of the disk is at $r_{\rm in}=16M$, and the outer edge of the disk is at $r=22M$. {\em Left Panel} -- string cloud parameter $a=0.2$ and {\em Right Panel} -- string cloud parameter $a=0.5$. The BH mass is taken as $M=1$.}
\label{fig6}
\end{center}

\par
Figure 4 shows the redshift distribution of the direct images. Owing to the presence of the BH, blueshift and redshift exist simultaneously in the direct image. As illustrated, an increase in the inclination angle $\theta_{0}$ results in the blueshift exceeding the gravitational redshift component. Additionally, a decrease in the string cloud parameter $a$ leads to an expansion of the region of the redshift distribution.

\par
Figures 5 and 6 depict the flux distributions of the direct and secondary images, respectively. It can be observed that when the inclination angle is small, the flux distribution is symmetric. As the inclination angle increases, the asymmetry of the flux distribution increases. Furthermore, the decrease in the string cloud parameter $a$ causes an expansion of the region of the flux distribution for the direct and secondary images.

\par
We generate a Schwarzschild string cloud BH image through a numerical simulation and compare the results with those obtained by the EHT. It is noteworthy that the image of the target BH bears a resemblance to the one portrayed in the Hollywood movie ``Interstellar.'' While the observation of the BH shadow is dependent on the EHT resolution, this basic numerical simulation provides a rough illustration of the EHT's capabilities.
\begin{center}
\includegraphics[width=5cm,height=4.2cm]{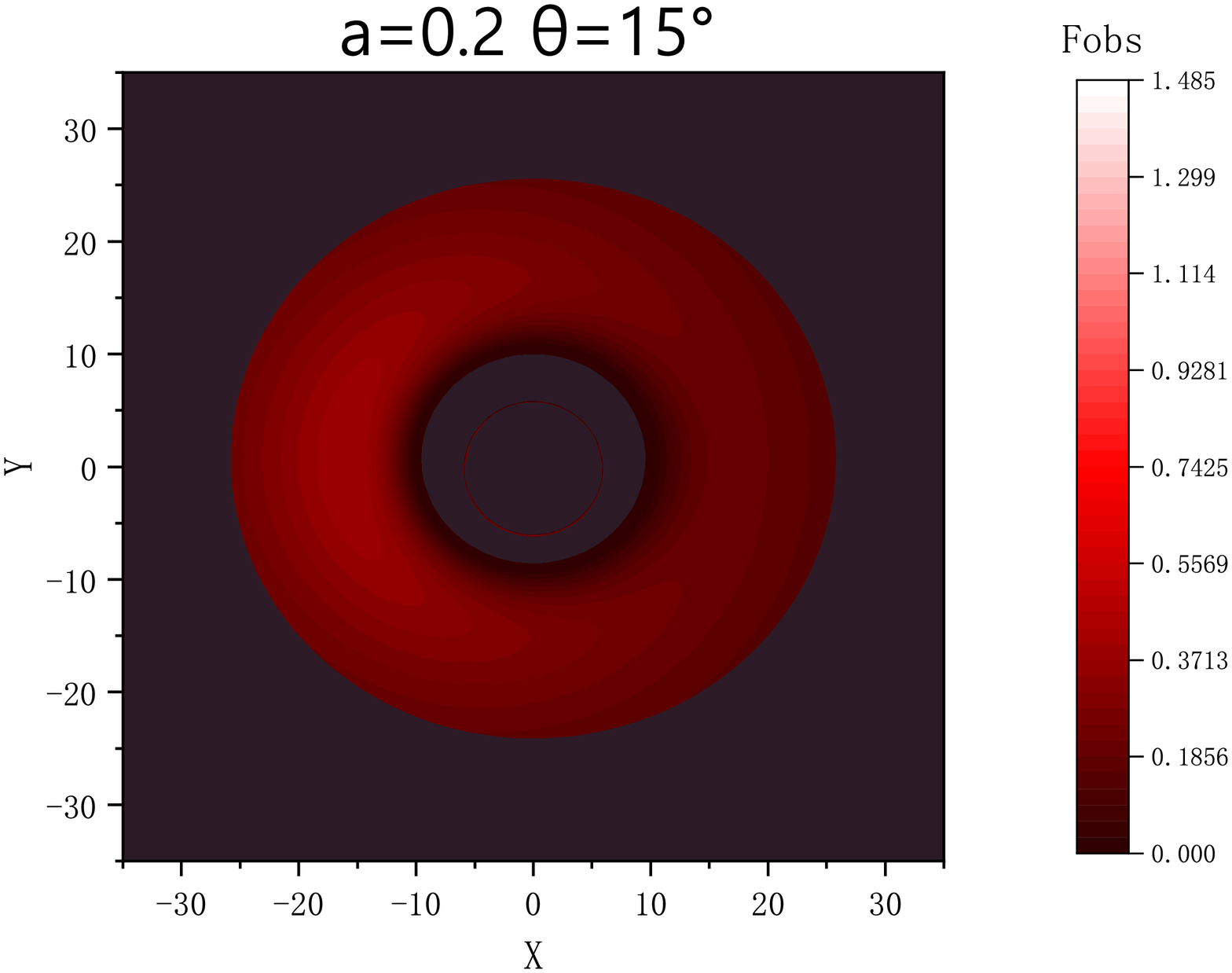}
\includegraphics[width=5cm,height=4.2cm]{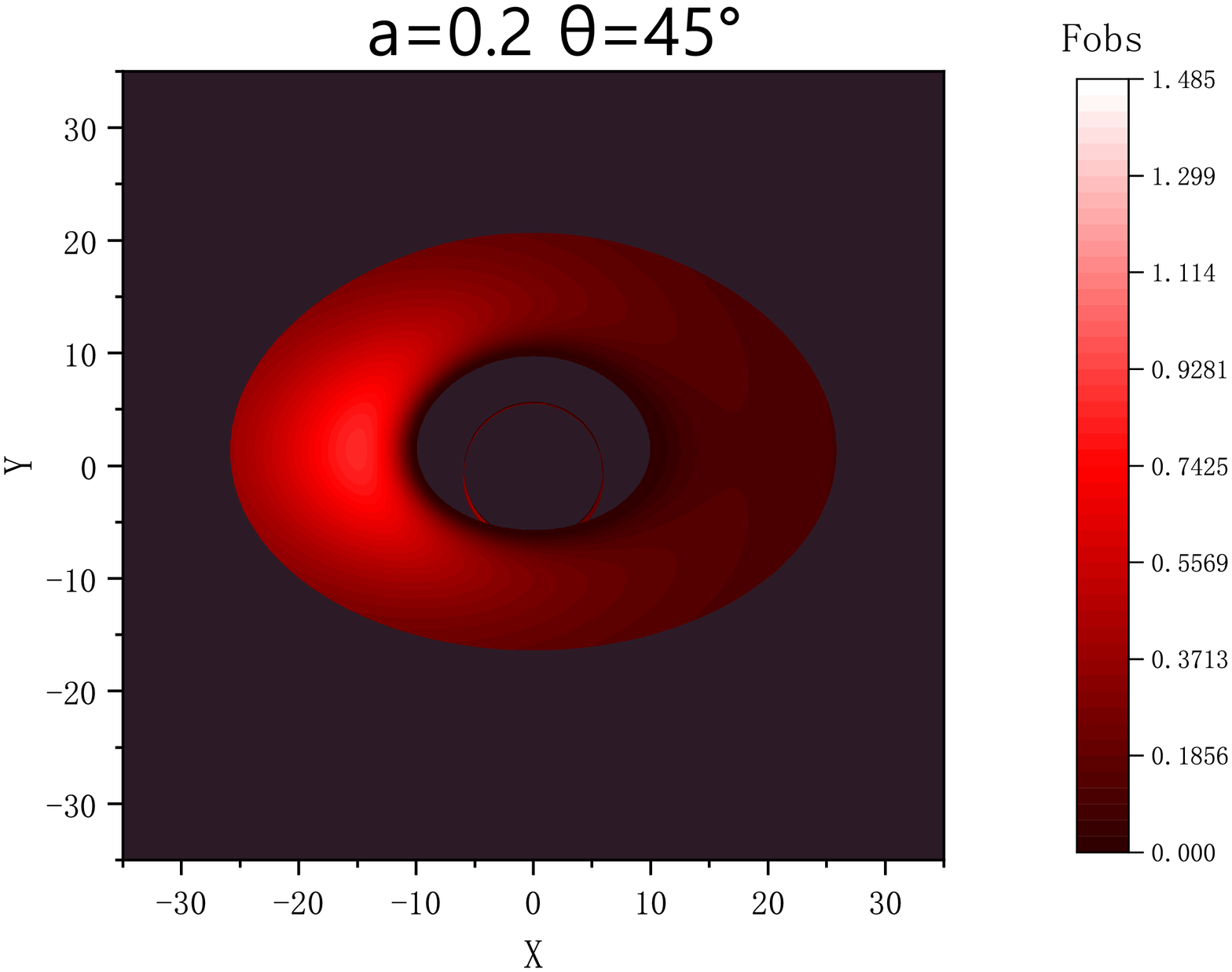}
\includegraphics[width=5cm,height=4.2cm]{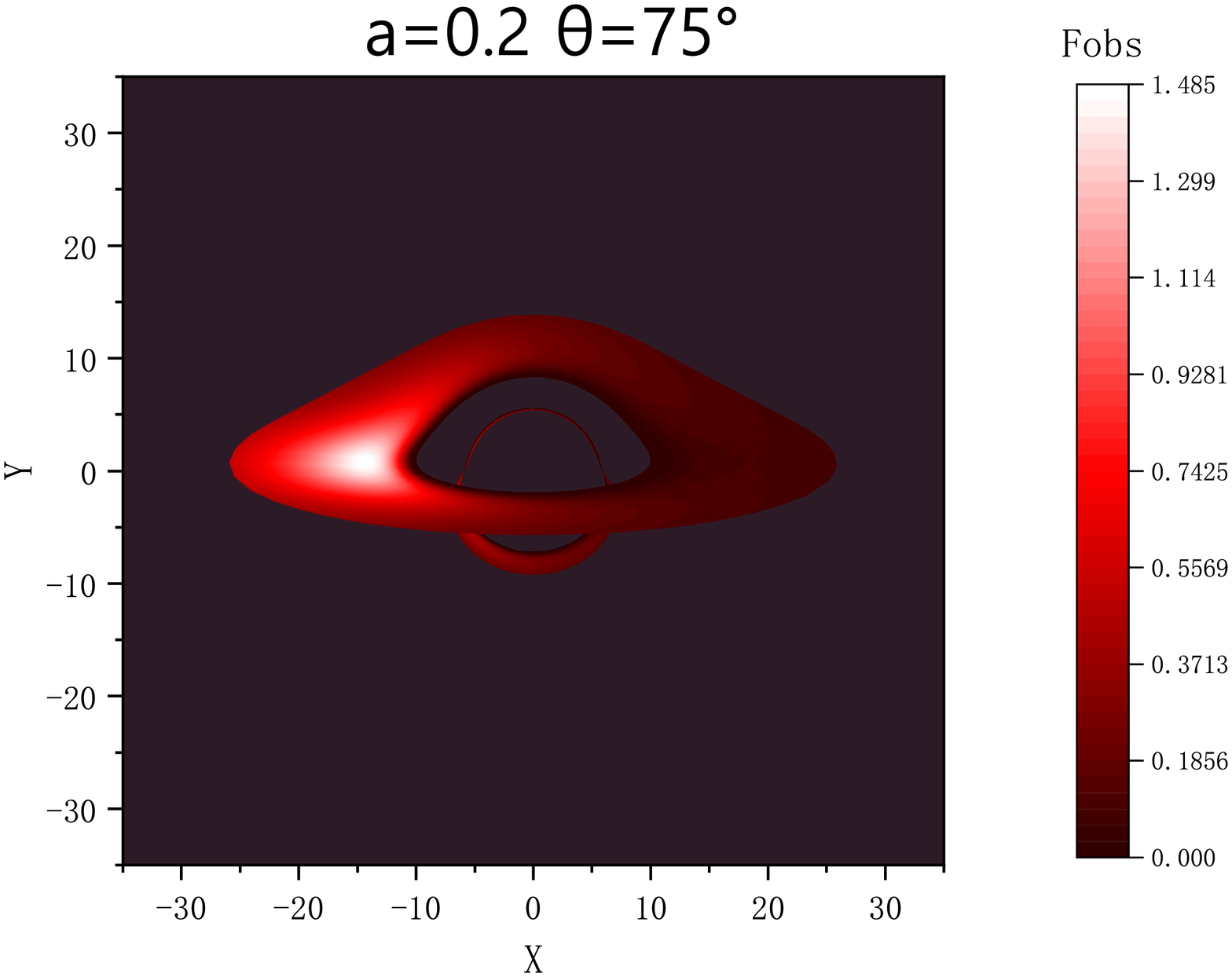}
\includegraphics[width=5cm,height=4.2cm]{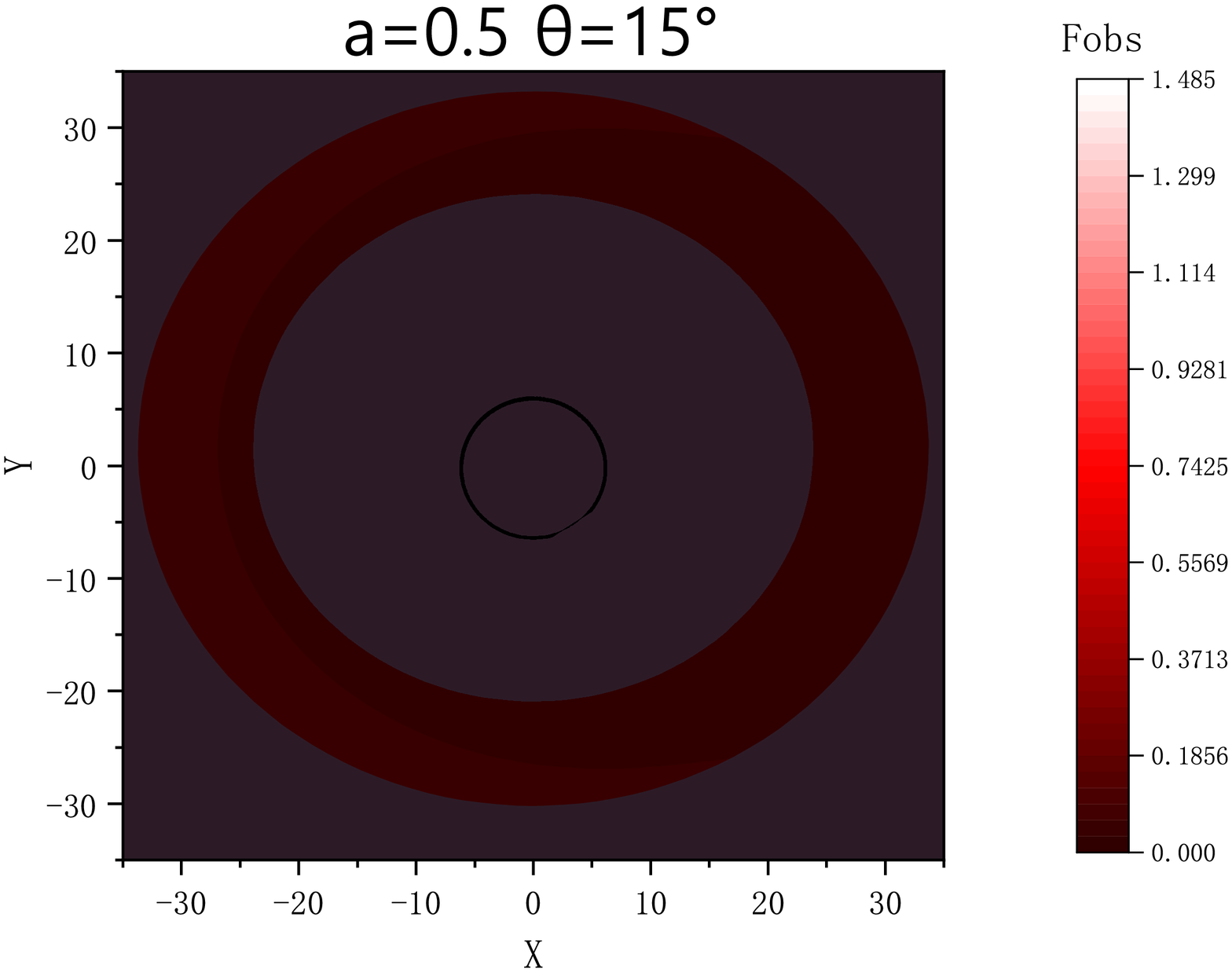}
\includegraphics[width=5cm,height=4.2cm]{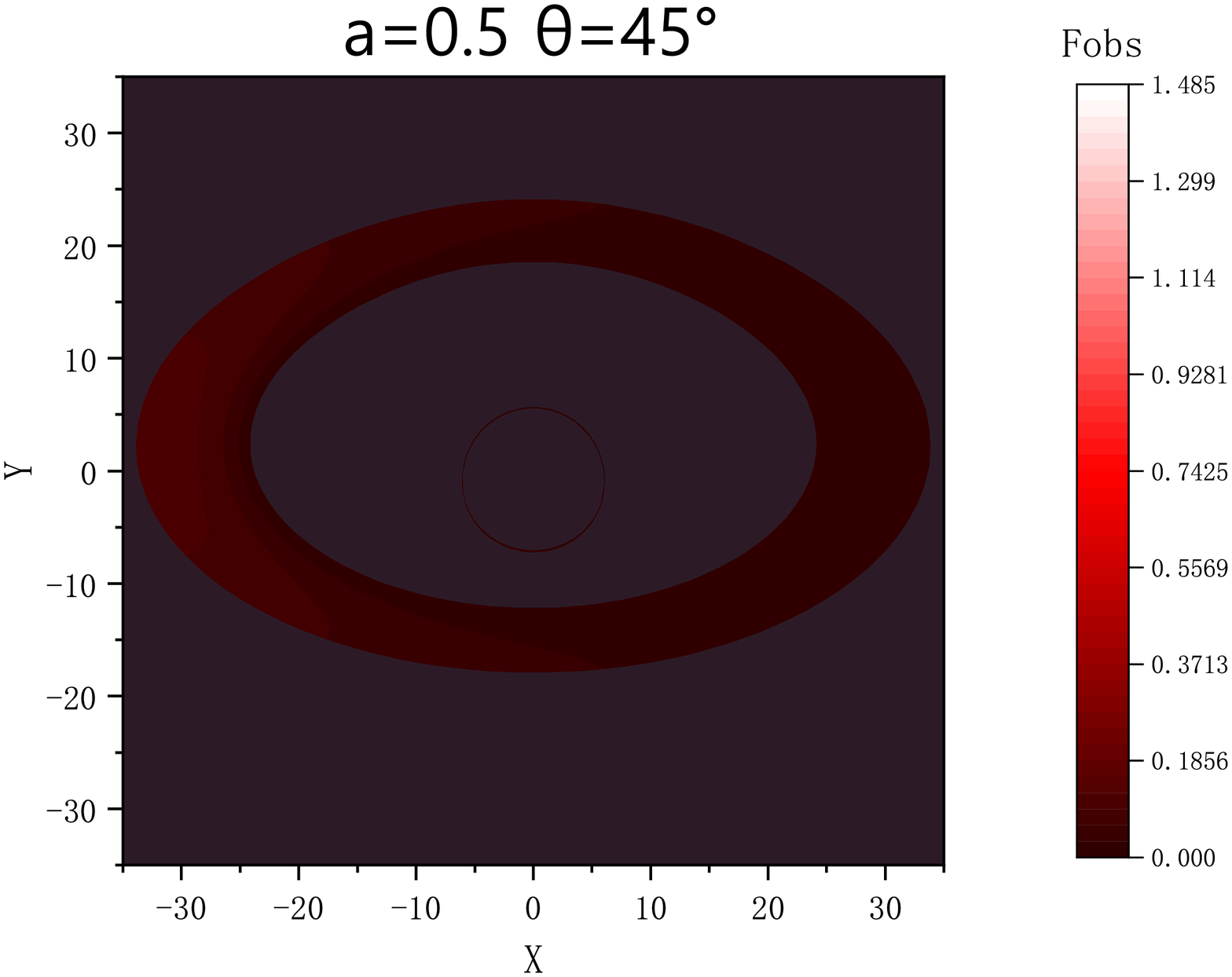}
\includegraphics[width=5cm,height=4.2cm]{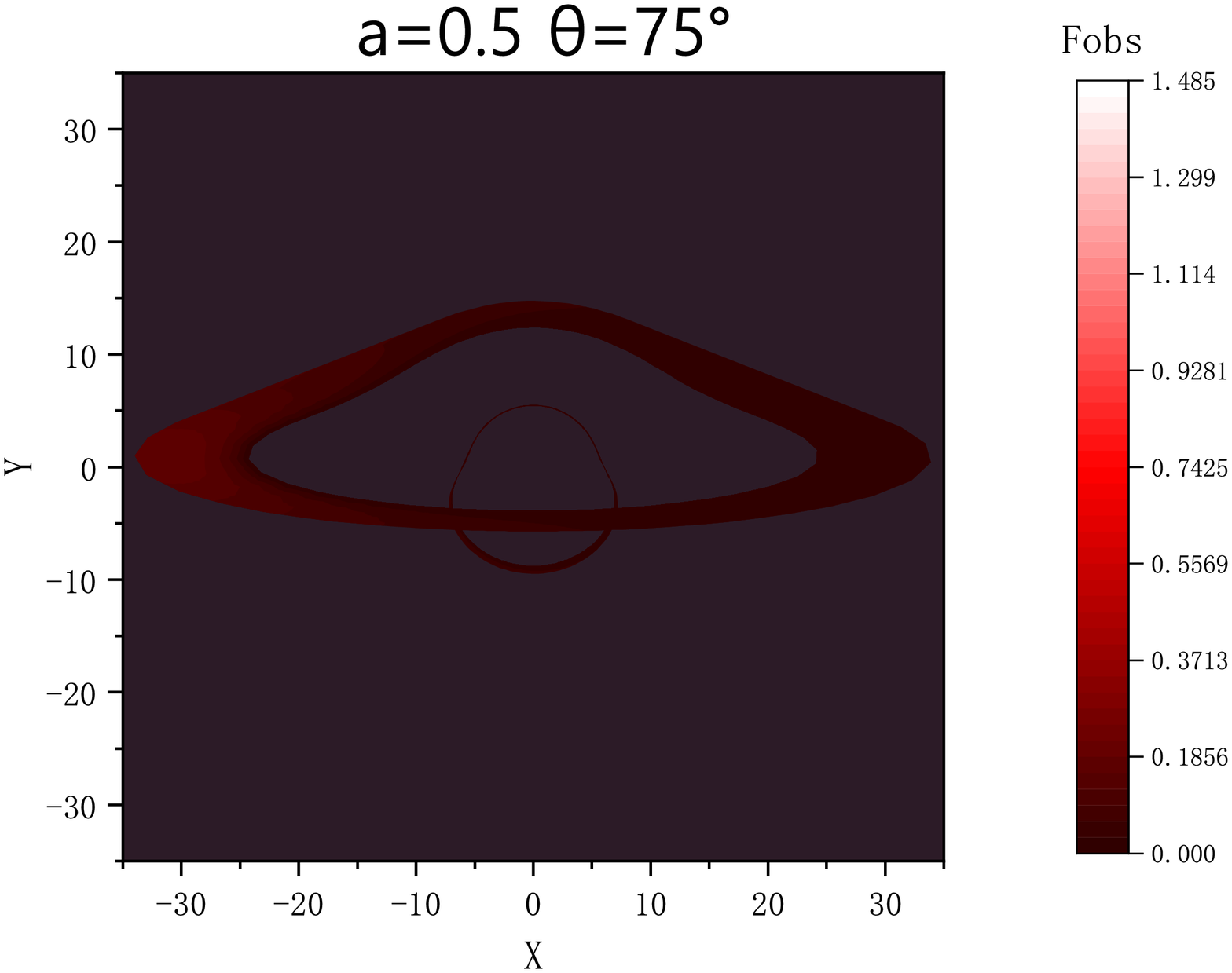}
\parbox[c]{15.0cm}{\footnotesize{\bf Fig~7.}  
Numerical simulation image of the Schwarzschild string cloud BH. The inner edge of the disk at $r_{\rm in}=16M$, and the outer edge of the disk is at $r=22M$. {\em Top Panel} -- string cloud parameter $a=0.2$ and {\em Bottom Panel} -- string cloud parameter $a=0.5$. The BH mass is taken as $M=1$.}
\label{fig7}
\end{center}

\section{Conclusions and Discussions}
\label{sec:5}
\par
In this analysis, we studied the optical appearance of the Schwarzschild BH in the context of a string cloud. We derived an effective potential function to calculate the total bending angle of light rays around the Schwarzschild string cloud BH. Using a ray-tracing code, we determined the trajectory of light rays around the BH. Our results showed that the string cloud parameter $a$ influences the size of the BH shadow, which can be highly curved for light rays that are close to the BH and have a large $a$ value. This suggests an increase in the light ray density for distant observers.

\par
Following the methodology in Ref. \cite{9}, we analyzed the direct and secondary images of the Schwarzschild BH surrounded by a string cloud. Our results show that the direct (secondary) image depends on the direction of photons emitted above (below) the equatorial plane. We observed that increasing the string cloud parameter $a$ value results in an increase in the direct image, which implies that the shadow radius expands outward when the string cloud size increases. However, the shape and size of the secondary images decrease for larger values of $a$. Moreover, we found that the inclination angle of the observer influences the separation between the direct and secondary images, with a more noticeable separation for larger inclination angles.

\par
We calculated the observed flux function of the Schwarzschild string cloud BH, taking into account gravitational redshift. Our analysis revealed that the direct images exhibit both blueshift and redshift, with the blueshift surpassing the gravitational redshift component as the inclination angle $\theta_{0}$ increases. Moreover, we observed that the redshift distribution expands with a reduction in the string cloud parameter $a$. Additionally, we examined the flux distributions of the direct and secondary images. Our results showed that the flux distribution is symmetric for smaller inclination angles. Furthermore, decreasing the string cloud parameter $a$ leads to an expansion of the region of the flux distribution for the direct images. We also performed a numerical simulation to obtain the image of the Schwarzschild string cloud BH, which provides a rough illustration of the EHT resolution.

\par
Comparing our findings with those of the EHT, we observed that our results are similar to the EHT results for small observation dip angles and exhibit a hat-like shape for large dip angles. As the resolution of the EHT continues to improve, it is possible that BH photographs similar to our results can be taken. Therefore, our work can serve as a theoretical guide for the enhancement of the EHT resolution. However, it is important to note that the asymmetry of bright rings in the EHT results is due to the fact that most astrophysical BHs are Kerr-like BHs. In our analysis, we did not include the accretion disk image of the Kerr BH, owing to the significant difference between the photon geodesics of a rotating BH and a spherically symmetric BH. Nevertheless, ongoing work on the image of the accretion disk of the Kerr BH is currently underway, and the next stage of our research will involve a more comprehensive study of the accretion disk of the rotating BH.

\section*{Acknowledgments}
This work is supported by the National Natural Science Foundation of China (Grant No. 11903025) and the Natural Science Foundation of Sichuan Province (2022NSFSC1833).

\clearpage

\end{CJK}
\end{document}